\documentclass{aa} 

\makeatletter
\let\makeLineNumber\@empty
\makeatother

\usepackage[colorlinks=true, linkcolor=blue, citecolor=blue, urlcolor=blue]{hyperref}
\usepackage{graphicx}
\usepackage{txfonts}
\usepackage{lipsum}
\usepackage{booktabs}
\usepackage{subcaption}         % necessary for continued figures, example in section 3
\usepackage{lscape}             % to rotate a single page table, example in appendix.
\usepackage{placeins}           % useful with \FloatBarrier, to keep 
\newcommand{\xmm}{{\it XMM-Newton}\xspace}
\newcommand{\gaia}{{\it Gaia}\xspace}
\newcommand{\nway}{\textsc{Nway\,}}
\begin{document}

	\title{X-ray luminous late-type giants: an overlooked population contributing to the Galactic ridge iron line emission}
%   \subtitle{Subtitle}

%%%%%%%%%%%%%%%%%%%%%%%%%%%%%%%%%%%%%%%%
% Please do not include ORCIDs next to author names.
% Only ORCIDs authenticated by individual authors in EDP Sciences editorial system will be taken into account.
% ORCIDs included here will be removed.
%%%%%%%%%%%%%%%%%%%%%%%%%%%%%%%%%%%%%%%%

   \author{Tong Bao
          \inst{1},
          Gabriele Ponti\inst{1,2,3},
          Xiao-jie Xu\inst{4},
          Mark R. Morris \inst{5},
          Benjamin Levin \inst{6},
          Kaya Mori \inst{6},
          Shifra Mandel \inst{6},
          Nicola Locatelli \inst{1},
          T. ~Mu\~noz-Darias \inst{7,8},
          J. ~Casares \inst{7,8},
          M.~A.~P.~ Torres \inst{7,8}
          }
   \institute{$^1$ INAF -- Osservatorio Astronomico di Brera, Via E. Bianchi 46, 23807 Merate, Italy
              \\\email{tong.bao@inaf.it}
              \\
         $^2$
             Max-Planck-Institut Extraterrestriche Physik,  GieAenbachstraAe1, 85748 Garching, Germany\\
         $^3$
          Como Lake Center for Astrophysics (CLAP), DiSAT, UniversitA degli Studi dellaInsubria, via Valleggio 11, I-22100 Como, Italy\\
         $^4$
          School of Astronomy and Space Science, Nanjing University, Nanjing 210046, China\\
         $^5$ Department of Physics and Astronomy, University of California, Los Angeles, CA, 90095-1547, USA\\
         $^6$ Columbia Astrophysics Laboratory, Columbia University, New York, NY 10027, USA\\
         $^7$ Instituto de Astrofisica de Canarias, E-38205 La Laguna, Tenerife, Spain\\
         $^8$ Departamento de Astrofisica, Universidad de La Laguna, E-38206 La Laguna, Tenerife, Spain\\
         }
   \date{Received April 30, 2026}

% \abstract{}{}{}{}{}
% 5 {} token are mandatory
 
  \abstract
  {The origin of the highly ionized iron emission (Fe~\textsc{xxv} at $6.7\,\mathrm{keV}$) characterizing the Galactic ridge X-ray emission (GRXE) remains a fundamental puzzle in high-energy astrophysics. Although the GRXE continuum was largely resolved into discrete populations of cataclysmic variables and coronally active stars, subsequent measurements revealed that these sources exhibit Fe~\textsc{xxv} equivalent widths significantly lower than that of the total GRXE, leaving the intense iron line emission unexplained.}
  {We aim to identify and characterize hard X-ray sources ($>2$ keV) within the XMM-Newton heritage survey of the inner Galactic disk that possess reliable Gaia counterparts. By analyzing their X-ray spectra and the presence of highly ionized iron emission lines, we seek to determine the physical nature of these hard X-ray emitters among the late-type giant population and evaluate their collective contribution to the Galactic X-ray background.}
   {We cross-correlated the XMM-Newton survey of the inner Galactic disk with Gaia DR3 astrometry. Sources located within the red giant branch of the color-magnitude diagram were selected for X-ray spectral analysis due to their significant hard X-ray emission. We derived their X-ray luminosities ($L_{\mathrm{X}}$) and spectral hardness ratios, with a particular focus on identifying signatures of extremely hot plasma, notably the Fe~\textsc{xxv} line complex at $\sim 6.7$ keV. Whenever possible, the optically variable nature of these sources was verified against established long-period variable identifications.}
{We identified 107 X-ray sources consistent with the long period variable populations in the Gaia color-magnitude diagram. These sources exhibit high X-ray luminosities ($L_{\mathrm{X}} \approx 10^{31}$-$10^{33} \mathrm{\,erg~s^{-1}}$), notably exceeding the typical saturation levels of single giants. Their X-ray spectra are significantly harder than those of quiescent stellar coronae, with plasma temperatures reaching up to $kT \approx 6$ keV added a prominent emission feature at $\sim 6.7$ keV.}
% conclusions heading (optional)
{The combination of high $L_{\mathrm{X}}$, hard X-ray spectra, and prominent $6.7 \text{ keV}$ Fe~\textsc{xxv} emission identifies these sources as a distinct population of accretion-powered binaries associated with late-type giants. Our analysis shows this population contributes $\sim$20\% of the total GRXE continuum and $\sim$40\% of its iron line emission.}

   \keywords{X-rays: stars -- binaries: symbiotic -- Galaxy: disk -- Galaxy: stellar content}

   \titlerunning{X-ray Luminous Late-Type Giants}
   \authorrunning{Bao et al.}
\maketitle

\section{Introduction}
The Milky Way is permeated by the Galactic ridge X-ray emission (GRXE) \citep{Worrall1983}, a vast glow characterized by a $10^8$ K optically thin thermal plasma spectrum with a prominent iron emission line at 6.7 keV \citep{Koyama1986}. 
Although initially interpreted as a hot, diffuse interstellar medium, high spatial resolution observations have successfully resolved the bulk of the GRXE continuum into discrete point sources \citep{Revnivtsev2009}, and suggest cataclysmic variables (CVs) and active binaries (ABs) as the primary constituents \citep{Hong2012,Nobukawa2016}. However, a fundamental discrepancy remains: these stellar populations exhibit Fe xxv equivalent widths significantly lower than the values measured for the total ridge emission \citep{Xu2016}. This deficit suggests the presence of either a truly diffuse plasma component or a hidden population of high-energy sources, either of which would fundamentally redefine our understanding of the Galaxy's energetic plasma distribution and its underlying heating history.
Furthermore, the outflow of hot plasma can be driven from this central region, potentially feeding the large-scale Galactic wind and transporting energy into the halo \citep{Ponti2015, Ponti2019}.

Among the potential contributors are symbiotic stars (SySts), which are wide binary systems consisting of a late-type giant star transferring mass to a compact companion, typically a white dwarf (WD) \citep{Allen1981, Allen1984, Belczynski2000}. Accretion onto the WD can power strong X-ray emission, making it detectable as a point source. The association of SySts with an old stellar population (late-type giants) makes them ideal candidates for a smoothly distributed population along the Galactic plane. 
However, the current census of SySts is likely significantly underestimated in X-ray population studies. As noted by \citet{Mukai2016}, SySts powered by accretion alone often exhibit weak optical emission lines, particularly in systems with low-luminosity donors or when observed via low-resolution spectroscopy. Consequently, current symbiotic star catalogs are heavily biased toward systems with prominent optical spectral features. This selection effect leaves a significant fraction of ``accretion-only'' SySts, which are often hard X-ray emitters \citep{Luna2013}, largely unaccounted for in existing surveys \citep{Mukai2017}.

Single late-type giants are generally not expected to be significant X-ray emitters. Despite possessing deep convective envelopes, these stars are typically slow rotators. The primary reason for this slow rotation is the dramatic expansion of the stellar radius during the post-main-sequence transition as a star swells by two orders of magnitude, the conservation of angular momentum dictates a corresponding decline in the surface rotation rate. Consequently, their magnetic dynamos are inefficient, failing to sustain the magnetically heated coronae required for hot plasma emission \citep{Linsky1979, Simon1982}.
Surveys conducted with the Einstein Observatory \citep{Ayres1981} and ROSAT \citep{Haisch1991, Hunsch1998B, Schroeder1998} have confirmed that most late-type giants are orders of magnitude less X-ray luminous than their main-sequence counterparts.

Nevertheless, early ROSAT observations identified X-ray emission from certain late-type giants, including "hybrid" stars that exhibit both hot coronal and cool-wind signatures \citep{Reimers1996}. Subsequent high-resolution studies, however, suggested that this flux often originates from unresolved active companions rather than the giants themselves \citep{Ayres2005}. 
 More recently, observations with XMM-Newton \citep{Ortiz2021} and the eROSITA all-sky survey \citep{Guerrero2024, Schmitt2024, Locatelli2025} have detected X-rays from M-type giants and asymptotic giant branch (AGB) stars. The reported luminosities, typically $L_{\rm X} \approx 10^{30}$ erg s$^{-1}$, significantly exceed the levels expected from standard coronal activity, named as ``X-AGB'' stars. There is now a general consensus that X-ray emission from these late-type giants, particularly when luminosities exceed a few times $10^{29}$ erg s$^{-1}$, serves as an indicator of binarity.

While the binary nature of these sources is well-established, the specific X-ray radiation mechanisms remain a subject of active investigation. The emission may stem from the active corona of a main-sequence companion or, more likely at higher luminosities, from accretion activity around a secondary star. At higher X-ray luminosity range ($\approx 10^{32}$ erg s$^{-1}$), these systems are typically identified as SySts where the giant donor transfers material to a WD via Roche-lobe overflow or wind accretion. This material is heated within an accretion disk and a boundary layer near the accretor, producing optically thin thermal X-ray emission that extends to several keV \citep{Luna2013, Lima2024}.
For sources in the lower luminosity range, X-ray emission may originate from accretion disks surrounding main-sequence companions, typically characterized by relatively soft spectra ($kT \lesssim 2$ keV; \citealt{Guerrero2024}). Furthermore, in binary systems close enough for mass transfer to occur, the red giant can be significantly spun up via tidal interactions with its companion. This process reinvigorates the magnetic dynamo and enhances coronal activity, a phenomenon observed even in advanced stages such as the AGB (e.g., V Hya; \citealt{Barnbaum1995}). Regardless of the specific companion type, the detection of hard X-ray emission from late-type giants serves as a robust indicator of binarity and interaction-driven evolution.

Our recent deep X-ray surveys of \xmm towards the inner Galactic disc have further characterized the populations contributing to the GRXE. In particular, \citet{Mondal2025} suggest that up to 50\% of the ridge emission in the 6.5--7 keV band can be accounted for by point sources at flux limits above $\sim 10^{-14}$ erg s$^{-1}$ cm$^{-2}$. 
Building upon these findings, this work reports a systematic search for X-ray luminous late-type giants across an extensive portion of the inner Galactic plane. By combining the high sensitivity of XMM-Newton with the unparalleled astrometric and photometric precision of Gaia DR3 \citep{GaiaDR3}, we isolated a robust sample of X-ray-luminous late-type giants. Our subsequent multi-wavelength analysis reveals that this population is likely to be a mixture of SySts and the ``X-AGB'' stars . We utilize this identified sample to provide a quantitative estimate of their contribution to the GRXE, offering new insights into the unresolved components of the Galactic ridge.

%%%%%%%%%%%%%%%%%%%%%%%%%%%%%%%%%%%%%%%%%%%%%%%%%%%%%%%%%%%%%%
\section{XMM-Newton survey and data preparation}
\label{sec:xdata}
This study is based on the XMM-Newton survey of the inner Galactic plane, covering a contiguous area of approximately 17x2 deg$^2$ ($-10\degr < l < 7\degr$ and $ -1\degr < b < 1\degr$). The average exposure is approximately 20 ks per tile.

\subsection{Source catalog}
We applied the XMM-Newton Serendipitous Source Catalog from Stacked Observations (4XMM-DR14s; \citealp{4XMMDR14s}). This catalog was constructed from simultaneous source detection on overlapping observations between 2001 and 2023, except for a 15\arcmin\ region about the Galactic Center, which was excluded from the processing for technical reasons. Of the 606 observations from our ongoing XMM-Newton Heritage Survey, 562 are included in this study. The remaining observations were either conducted after 2024 or lie within the excluded Galactic Center region. We applied rigorous quality cuts as follows:  
\begin{itemize}
  \item STACK\_FLAG $\leq$ 1 (minimizes stacked detection artifacts)  
  \item EP\_DET\_ML $\geq$ 6 (false-positive probability $<0.1\%$)  
  \item EXTENT = 0 (include point sources only)  
\end{itemize}
The final sample comprises 15655 sources falling within the footprint of our {\it XMM-Newton} Heritage survey, 9533 of which were observed in more than one observation. 
The observation data files were processed using the \xmm Science Analysis System (SAS, v22.1.0\footnote{https://www.cosmos.esa.int/web/xmm-newton/sas-release-notes-2210}). 
We used the task \emph{evselect} to construct a high-energy background light curve (energy between 10 and 12 keV for EPIC-pn \citep{struder2001} and above 10 keV for EPIC-MOS1/MOS2 \citep{turner2001}) by selecting only PATTERN==0. The background light curve was used to filter high-background flaring activity and to thereby identify good time intervals.

\subsection{Cross-matching \xmm and \gaia Sources}

We used data from the third data release of the \textit{Gaia} mission (DR3). The key parameters for our analysis are the $G$, $G_{\mathrm{BP}}$, and $G_{\mathrm{RP}}$ magnitudes, together with the parallax measurements, which are essential for estimating distances and luminosities.  
Since the stellar density varies significantly across the sky, we divided our \xmm\ survey coverage into longitudinal bins of $1^{\circ}$ width and extracted all \textit{Gaia}~DR3 sources within each bin to construct a subset of the source catalogue.

We performed a positional cross-match between our XMM-Newton source catalog and the Gaia DR3 catalog using the \nway algorithm \citep{Salvato2018} for each subset.
\nway has been developed for identifying the multi-wavelength counterparts to X-ray sources in multiple catalogues using a multidimensional parameter space (e.g. position and positional uncertainty, density of sources, magnitudes, colours, variability, morphology, etc.) in a Bayesian framework.
In short, \nway first computes for each source in the Gaia DR3 catalogue the Bayes factor considering only distance from the X-ray source, positional uncertainties and number densities. 
Then, each Gaia source is associated with the probability $p_i$ of being the true counterpart to a specific X-ray detection. In addition, for each X-ray detection, \nway provides the probability, $p_{\text{any}}$, that any of the Gaia sources is the actual counterpart. The higher the value of $p_{\text{any}}$, the lower the probability of a chance association.  

This process yielded 2752 unique X-ray sources with a reliable Gaia counterpart, with a false selection rate lower than 1\%.
The false selection rate  was calibrated by simulating chance alignments. This involved creating a fake catalogue by positionally shifting the source catalogue by a distance much larger than the positional errors and running the \nway matching algorithm with identical settings. The resulting output was then used to determine the $p_{\text{any}}$ cut-off limit corresponding to the required false selection rate.

\section{Gaia Hertzsprung--Russell diagram with X-ray properties}
\label{sec:HRX}
We construct the Hertzsprung--Russell diagram using extinction-corrected absolute $G$-band magnitudes ($M_{\rm G}$) and $\rm BP-RP$ colours.
The absolute magnitudes and intrinsic colours were derived by applying corrections for interstellar extinction and adopting reliable distance measurements.
For distances, we prioritized the photometric distance estimates, $\texttt{distance\_gspphot}$, provided in the \gaia\,DR3 catalogue. Where this measurement was unavailable, we adopted the inverse parallax, $1/\pi$, only if the precision satisfied $\texttt{parallax\_over\_error} > 3.0$.
The extinction correction utilised the three-dimensional dust map from \citet{Lallement2019}. The extinction coefficients necessary to convert the colour excess $\rm E(B-V)$ to the respective \gaia\,bands were adopted from \citet{Wang2019} using the following relations:
$\rm A_{\rm V} = 3.16 \times E (B-V)$,
$\rm A_{\rm G} = 0.789 \times A_V$,
$\rm A_{\rm BP} = 1.002 \times A_V$,  
$\rm A_{\rm RP} = 0.589 \times A_V$.

\begin{figure}[h!]
  \centering
  \includegraphics[width=0.99\hsize]{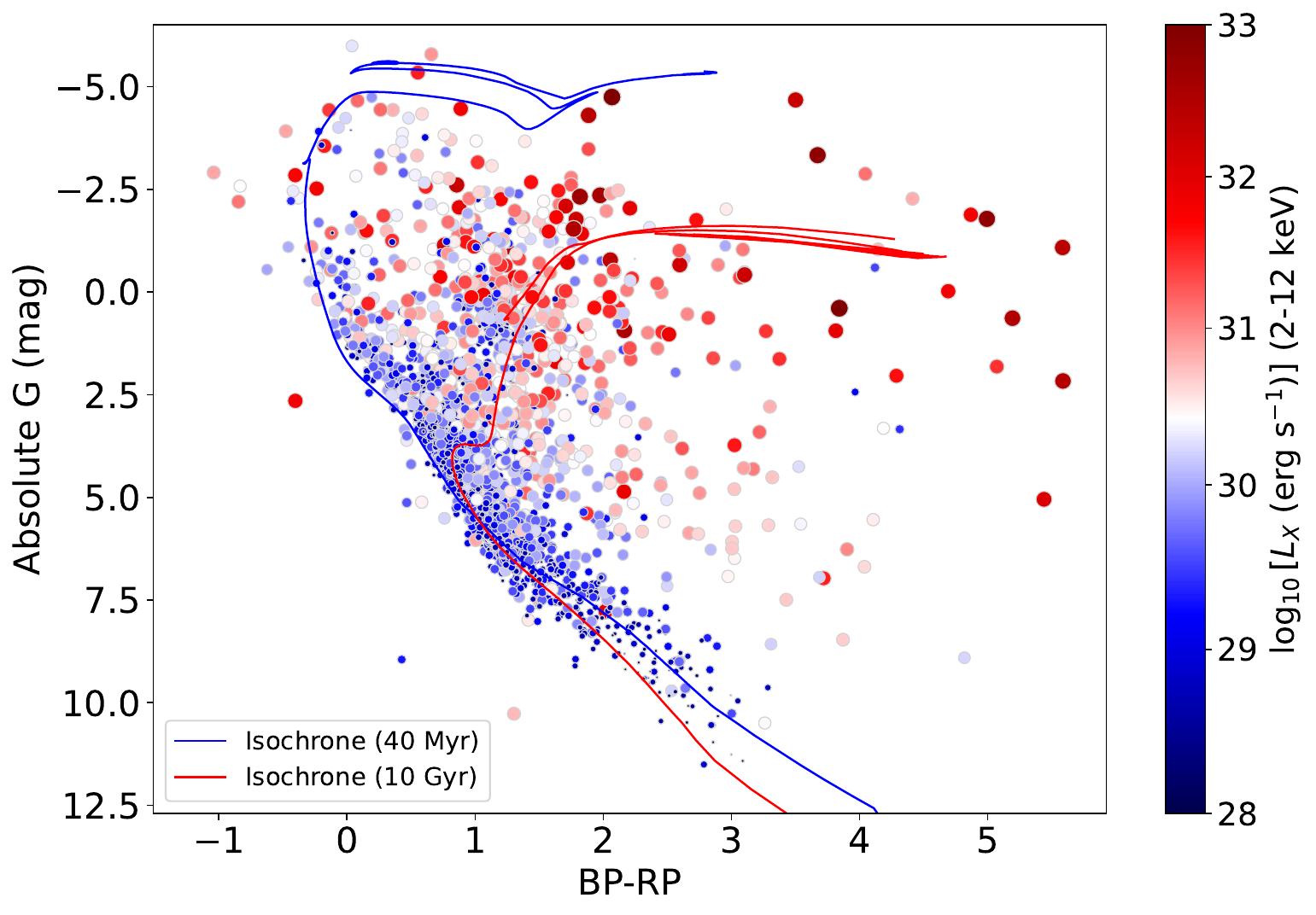}
  \caption{Hard X-ray luminosity in \gaia Hertzsprung-Russell diagram for sources with distance estimates. The x-axis shows the \gaia BP-RP color, and the y-axis shows the absolute $G$-band magnitude. Both the color bar and the symbol size indicate the absorbed 2--12~keV X-ray luminosity. Overplotted are PARSEC isochrones for stellar populations of 40 Myr (blue) and 10 Gyr (red).}

  \label{fig:HR_LX}
\end{figure}

Fig.~\ref{fig:HR_LX} displays the resulting diagram, where the colour scale encodes the relatively hard X-ray luminosity ($L_{\rm X}$) in the 2.0--12.0 keV band, which is calculated using the adopted \gaia\,distance and the observed X-ray flux.  For stellar context, we overlay PARSEC stellar isochrones corresponding to ages of 40\,Myr and 10\,Gyr, representing characteristic young and old stellar populations, respectively \citep{Bressan2012}.

\begin{figure}[htbp]
  \centering
  \includegraphics[width=0.99\hsize]{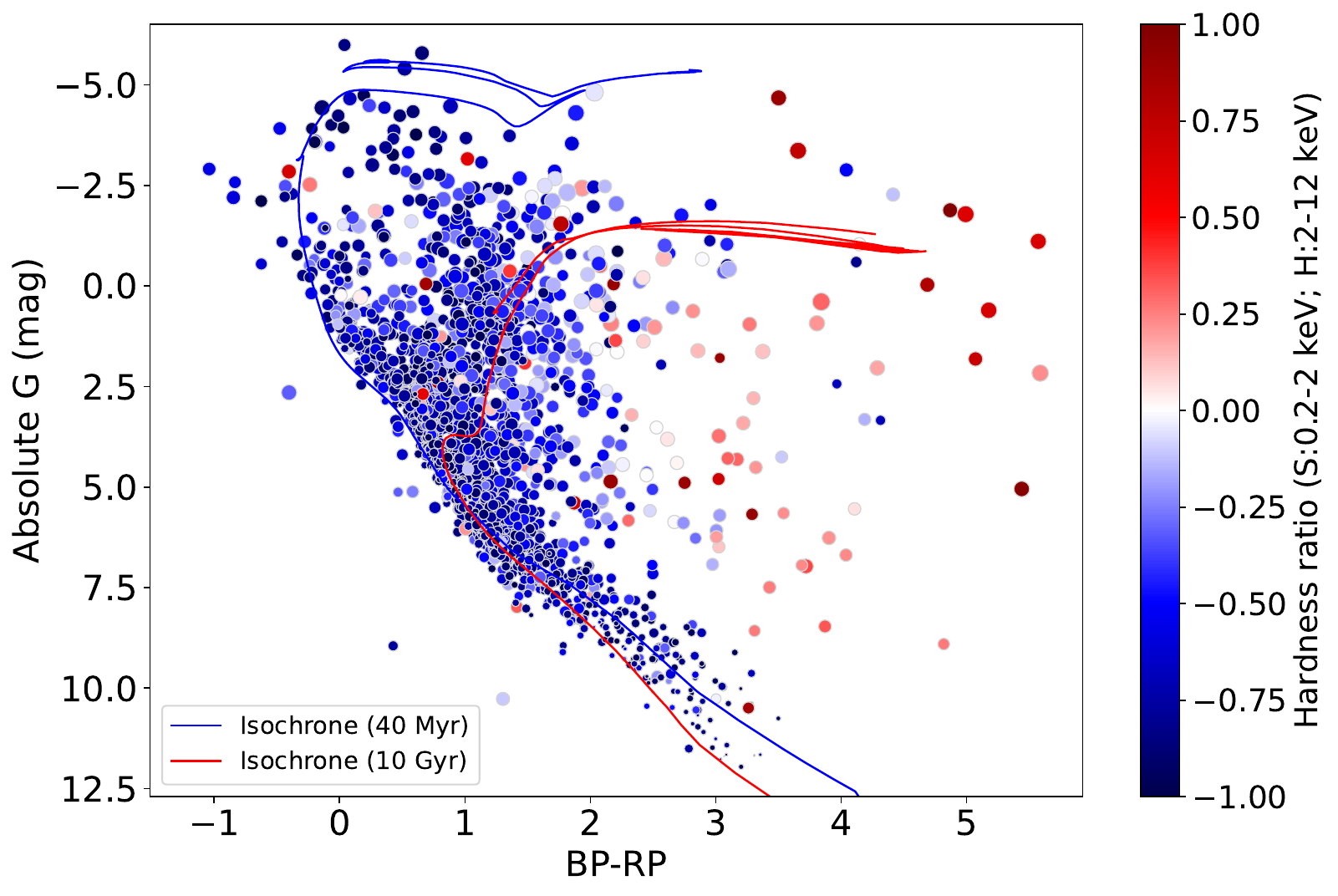}
  \caption{X-ray HR in \gaia Hertzsprung-Russell diagram for sources with distance estimates, using the same labels defined in Fig.~\ref{fig:HR_LX}. The color bar indicates the X-ray HR and the symbol size indicates the absorbed 2--12~keV X-ray luminosity. }
  \label{fig:HR_HDR}
\end{figure}

For supplementary characterisation, we also plot the \gaia\,magnitudes against the X-ray hardness ratio (HR) in Fig.~\ref{fig:HR_HDR}, defined as:
\begin{equation}
\mathrm{HR} = \frac{\rm H - S}{\rm H + S},
\end{equation}
where $S$ and $H$ represent the observed count rates in the soft (0.2--2~keV) and hard (2--12~keV) energy bands, respectively.

Inspection of Fig.~\ref{fig:HR_LX} and Fig.~\ref{fig:HR_HDR} reveals a clear dichotomy in the X-ray properties of the stellar populations. Sources located near the main sequence exhibit soft X-ray spectra and relatively lower X-ray luminosities. The stellar population situated to the right of the main-sequence, which is primarily expected to consist of late-type giants, is characterised by high X-ray luminosities and harder X-ray spectra.
The subsample of soft X-ray emitters detected by eROSITA (specifically eRASS1) in the same sky coverage has been discussed in \citet{Tong2025}, where two SySt candidates are identified due to their red, bright optical counterparts. In contrast, our XMM source sample reveals numerous sources in the late-type giants region with hard X-ray spectra, enabled by the superior sensitivity of our XMM heritage survey and the resulting larger source sample.
Both their location on the \gaia\ diagram and their hard X-ray luminosities indicate that these sources represent a population exclusively detected in \xmm and are markedly different from the coronal sources dominating the eRASS1 catalogue. 

\begin{figure}[h!]
  \centering
  \includegraphics[width=0.99\hsize]{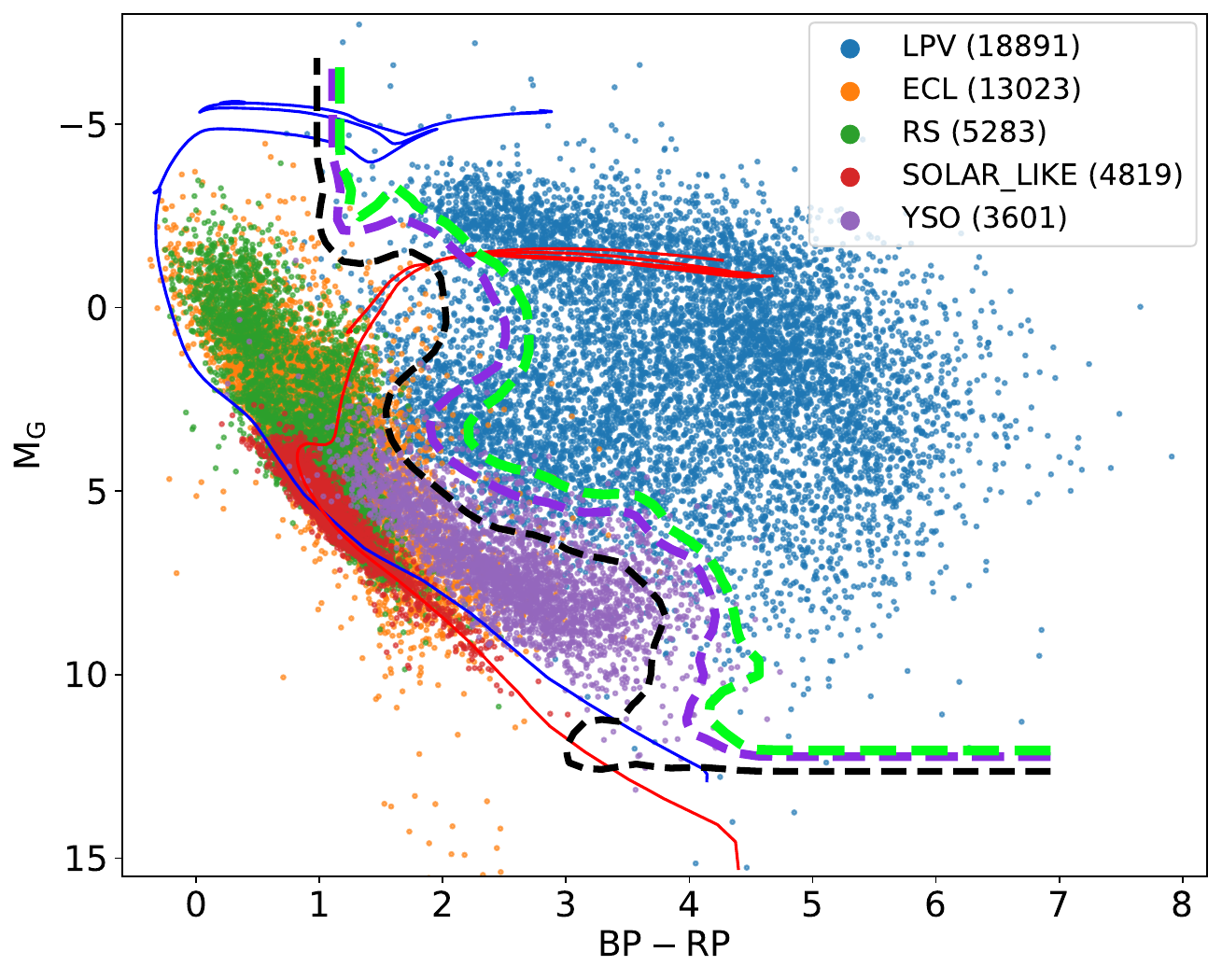}
  \caption{
    \gaia Hertzsprung-Russell diagram for sources with distance estimates. Absolute G-band magnitude is plotted against Gaia BP-RP color. Five source classes are color-coded as indicated in the legend. The blue, purple, and green dashed lines denote the purity contours for LPVs at confidence levels of 50\%, 80\%, and 90\%, respectively.}
  \label{fig:HR_contour}
\end{figure}

\section{Identification of X-ray luminous late-type giants}
The stellar population positioned redward of the main sequence in the \gaia color-magnitude diagram (CMD) is expected to be dominated by late-type giants. 
To precisely quantify the population of X-ray luminous giants motivated by the spectral dichotomy discussed in Sect.~\ref{sec:HRX}, we leverage the classifications for variable sources provided in the \gaia DR3 catalog.
By utilizing these distinct stellar classifications, we can define a robust confidence region in the CMD to isolate late-type giants from other contaminating populations.
The following subsections detail the selection process and the resulting connection between the observed X-ray properties and stellar evolutionary stages (as illustrated in Figs.~\ref{fig:HR_LX}--\ref{fig:HR_HDR}).
\subsection{Gaia DR3 classification of variable sources}
\label{subsec:gaiav}
\citet{gaiaclass} employed supervised machine learning techniques, specifically eXtreme Gradient Boosting and Random Forest, to classify variable objects into 25 distinct categories based on photometric time-series data in the $G$, $G_{\rm BP}$, and $G_{\rm RP}$ bands.
We extracted the classified sample within our study's footprint, focusing on the inner Galactic disk ($-10\degr < l < 7\degr, |b| < 1\degr$).
After applying the extinction corrections detailed in Sect.~\ref{sec:HRX}, we reconstructed the \gaia CMD for this classified sample (Fig.~\ref{fig:HR_contour}). The distributions of the five primary variable star classes are shown in Fig.~\ref{fig:HR_contour}, where long-period variables (LPVs) serve as the characteristic representatives of late-type giants. The other four main classes are eclipsing binary (ECL), RS Canum Venaticorum variable (RS), solar-like star (SOLAR\_LIKE), and young stellar object (YSO).
To define a robust selection region for the late-type giant population, we calculated the purity contours of the LPVs. We define purity as the fraction of LPVs relative to the total number of classified sources within each pixel of the CMD. We adopted the 80\% purity contour (represented by the thick purple dashed line in Fig.~\ref{fig:HR_contour}) as our primary selection boundary.

\subsection{X-ray and optical views of late-type giants}
%% Exclude the one with XMMidx 5620!!! 3055472010120010 srcid, is a HMXB 
LPVs represent a class of cool, luminous, pulsating evolved stars, primarily giants and supergiants with pulsation periods on the order of several hundred days. 
Based on the classification by \citet{gaiaclass}, our X-ray-detected sample includes 85 solar-like stars, 323 RS systems, 137 ECLs, 24 YSOs, and 49 LPVs. Additionally, 14 of the 49 LPVs lacked reliable parallax or distance measurements and were omitted from the luminosity-based analysis.

\begin{figure}[h!]
  \centering
  \includegraphics[width=1.0\hsize]{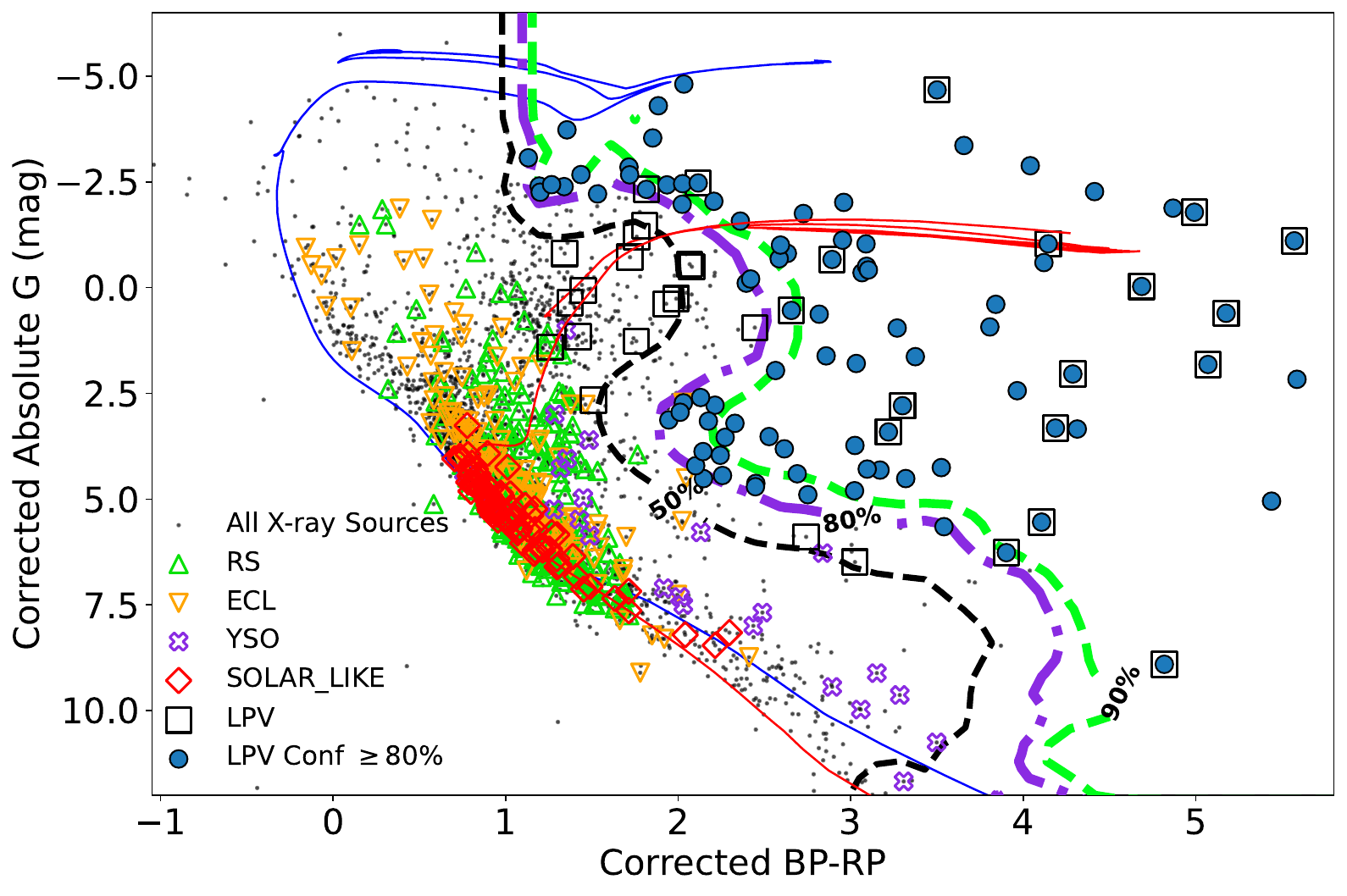}
  \caption{
    \gaia Hertzsprung-Russell diagram for various classes of variable sources. Sources located within the 80\% purity contour for LPVs are highlighted as blue circles and classified as LPV candidates. Other classes are represented using the color scheme defined in Fig.~\ref{fig:HR_contour}.
    }
  \label{fig:CMD_gaiaclass}
\end{figure}

In the \gaia CMD shown in Fig.~\ref{fig:CMD_gaiaclass}, solar-like sources are concentrated along the main sequence, while YSOs are located toward the lower-right, consistent with their pre-main-sequence nature. ECLs and RS systems occupy overlapping regions and extend toward the giant branch. This suggests that many ECLs in our sample may be active binaries physically similar to RS systems.

LPVs dominate the extreme redward region of the diagram, as expected for cool giants.
By applying the selection criteria defined in Sect.~\ref{subsec:gaiav}, we selected sources located within the LPV confidence region ($\text{purity} > 80\%$) alongside those already classified as LPVs by \citet{gaiaclass}. This procedure yielded 72 newly identified LPV candidates that share the same locus in the \gaia CMD as the known LPV population.

\begin{figure}[h!]
  \centering
  \includegraphics[width=1.0\hsize]{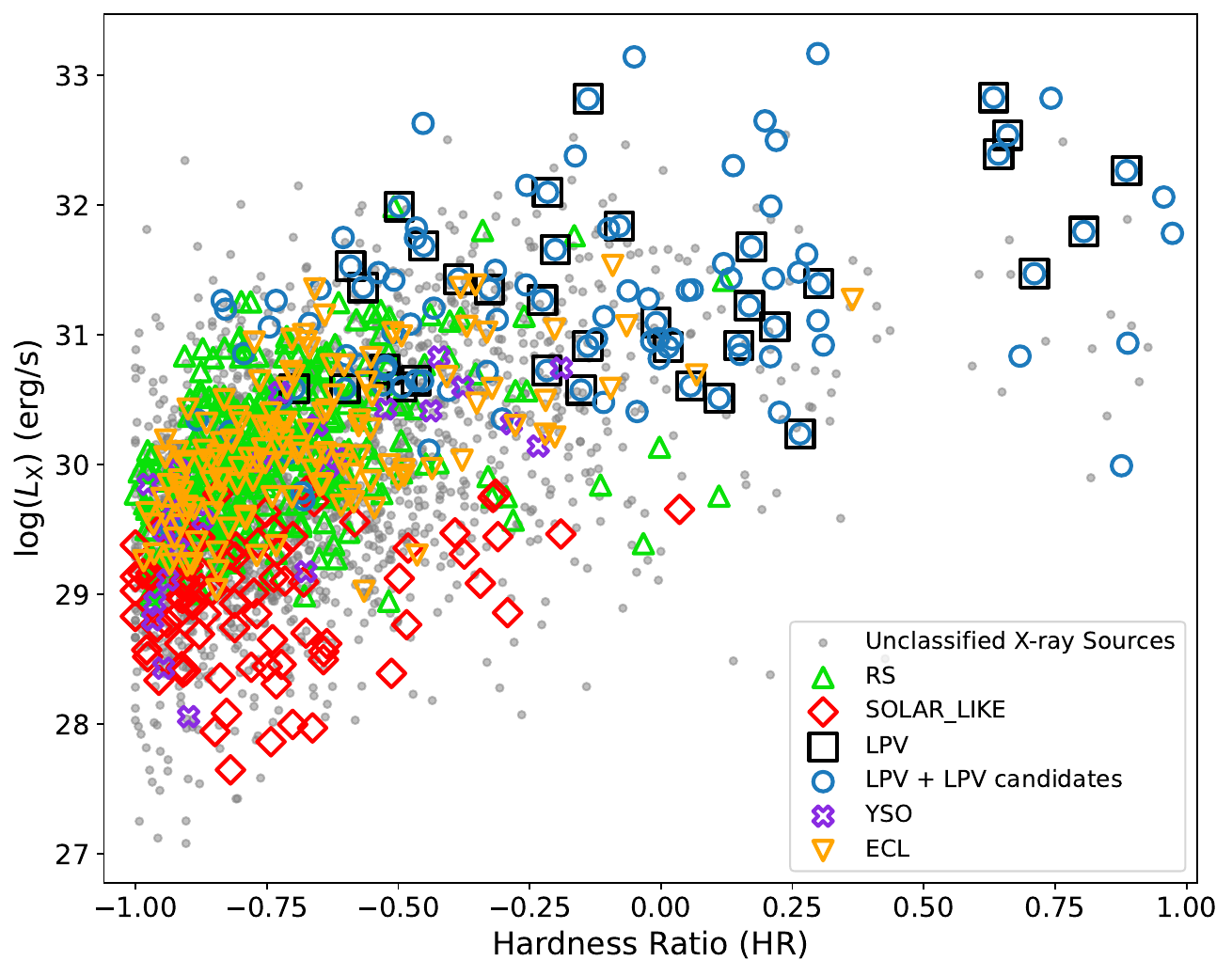}
  \caption{X-ray luminosity versus HR. Data points are categorized using the same labels and color scheme defined in Fig.~\ref{fig:HR_contour} and ~\ref{fig:CMD_gaiaclass}.}
  \label{fig:HDRLx}
\end{figure}

Fig.~\ref{fig:HDRLx} displays the X-ray hardness ratio as a function of the hard X-ray luminosity ($L_{\rm X}$) for the classified sources. 
As expected, solar-like stars, predominantly single active stars, exhibit soft spectra and low luminosities. 
In contrast, both the classified LPVs and the newly identified candidates show significantly harder X-ray spectra and elevated luminosities, typically ranging from $10^{31}$ to $10^{33}~\mathrm{erg~s^{-1}}$. The consistency between their X-ray and optical properties strongly suggests that this sample is representative of the X-ray-luminous late-type giant population. A detailed investigation into their X-ray spectral characteristics is presented in the following section.

\section{X-ray spectral analysis }
\subsection{X-ray spectra for different \gaia source classes}
\begin{figure*}[h]
  \centering
  \includegraphics[width=1.0\hsize]{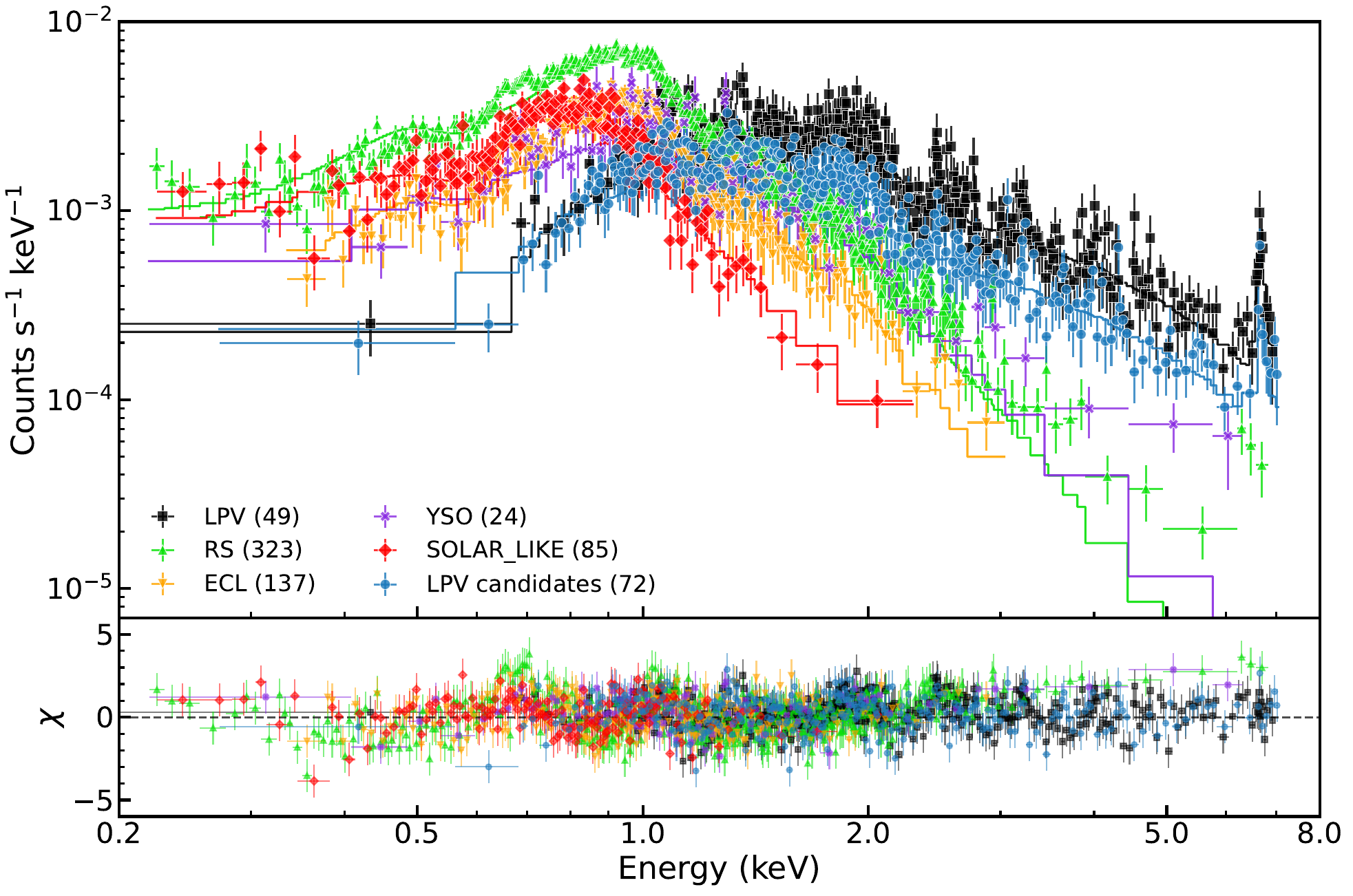}
  \caption{Stacked X-ray spectra of various source classes identified in \citet{gaiaclass}, along with the LPV candidates. The integer values provided in the legend indicate the number of individual sources within each respective class. For consistency, data points are categorized according to the same labeling and color scheme established in Fig~\ref{fig:CMD_gaiaclass}.}
  \label{fig:specfive}
\end{figure*}

With the well-classified \gaia sample, we stacked the X-ray spectra of each major source class identified in our X-ray source list, as shown in Fig.~\ref{fig:specfive}. 

For each source, the source and background spectra were extracted using the SAS task \emph{evselect}. 
Events were selected with \texttt{PATTERN}~$\leq$~4 for the EPIC-pn detector and \texttt{PATTERN}~$\leq$~12 for the MOS1 and MOS2 detectors.  
The source spectrum was extracted from a circular region with a radius of 25\arcsec\ centered on the source position, while the background was taken from an annular region centered at the same position, with inner and outer radii of 50\arcsec\ and 100\arcsec, respectively.  
Any nearby X-ray sources falling within the background region were excluded by masking circular regions of radius 25\arcsec.  
Ancillary response files (ARFs) and redistribution matrix files (RMFs) were generated using standard SAS procedures.  
The spectra from individual observations were then combined to produce a single, exposure-weighted spectrum for each source class using the task \texttt{epicspeccombine}.  
The combined spectra were grouped to achieve a signal-to-noise ratio (S/N) greater than two per energy bin.

Since most of these sources are expected to exhibit emission from optically thin thermal plasma, we adopted a phenomenological approach to spectral modeling using the Python interface to \texttt{XSPEC} \citep{gordon2021}.
Each stacked spectrum was fitted with an absorbed thermal plasma model (\texttt{tbabs}$\times$\texttt{apec}), providing a representative plasma temperature for each class.
The best-fitting models and parameters are summarized in Tab.~\ref{tab:t7}.
Their X-ray spectral plots are presented in Fig.~\ref{fig:specfive}.

The X-ray spectra of confirmed LPVs and LPV candidates are consistent with each other given the uncertainties, both featuring high plasma temperatures in the range of $kT \sim 5\text{--}6 \text{ keV}$. A prominent Fe~\textsc{xxv} emission line at $6.7 \text{ keV}$ is clearly detected, serving as a signature of highly ionized, thermal plasma. 
To quantify this, we measured the equivalent width (EW) of the $6.7 \text{ keV}$ iron line by setting the abundance to zero and incorporating three independent Gaussian components to fit the 6.4, 6.7, and 7.0~keV lines simultaneously.
Furthermore, we conducted a detailed spectral analysis of the less-luminous subsample to assess the potential impact of YSO contamination introduced during the selection process (see details in Appendix \ref{app:spectra}). The spectral congruity between the luminosity-selected subgroups reinforces the hypothesis that the observed high-temperature emission is an intrinsic property of the candidate population rather than contamination from other source classes.

Other source classes exhibit spectral temperatures broadly consistent with expectations from their known stellar activity levels.
Solar-like sources show relatively soft X-ray spectra with $kT \sim 0.7 \rm~keV$, consistent with results from the eRASS1 sample \citep{Tong2025,Locatelli2025,Zheng2026,Ponti2026}.
YSOs are characterized by higher plasma temperatures of about $kT \sim 1.3 \rm~keV$. Their X-ray emission originates from a combination of intense magnetic reconnection events in powerful coronae and, in some cases, additional heating in accretion shocks where material from the circumstellar disk impacts the stellar surface.
These processes make YSOs considerably hotter than magnetically active single stars \citep[e.g.,][]{Preibisch2005, Getman2005, Gudel2007}.
RS and ECL systems, dominated by tidally locked active binaries, display even higher coronal temperatures and, in some cases, an Fe-line excess near $6.7 \rm~keV$ contributed by a few particularly bright systems. 
Such behavior is consistent with strong magnetic activity sustained by rapid rotation and enhanced dynamo efficiency in close binary configurations.
The X-ray spectra of these active binaries (especially for RS) exhibit a hard X-ray excess beyond $2 \text{ keV}$ that cannot be adequately reproduced by a single-temperature plasma model, necessitating multi-thermal components for a physically consistent fit.

%\renewcommand{\arraystretch}{1.5}
%\begin{table}[h!]
%\centering
%\stepcounter{table}
%\label{tab:t7}
%\caption[]{}
\begin{table}[h!]
\centering
\renewcommand{\arraystretch}{1.5}
%\refstepcounter{table} 
\caption{Best-fit parameters for different source class.}
\label{tab:t7}
\begin{tabular}{lcccc}  
\hline\hline
Class &  $N_{\rm H}$ & kT & $\chi^2/$d.o.f & EW 6.7  \\
 &  $\rm (10^{21} cm^{-2})$ & (keV) &  & (keV)  \\
\hline
Solar-like  &  $0.62_{-0.36}^{+0.46}$ & $0.68_{-0.03}^{+0.04}$ & 1.05 (195) & / \\
RS   &  $1.68_{-0.06}^{+0.06}$ & $0.94_{-0.01}^{+0.01}$   & 1.14 (489) & / \\
ECL  & $1.89_{-0.14}^{+0.14}$  & $0.94_{-0.02}^{+0.02}$   & 1.07 (304) & / \\
YSO  & $2.13_{-0.39}^{+0.41}$   & $1.26_{-0.15}^{+0.31}$  & 1.00 (175) & / \\
LPV   & $5.54_{-0.58}^{+0.56}$ & $5.80_{-0.78}^{+1.26}$   & 0.90 (568) & $0.82_{-0.15}^{+0.15}$ \\

LPV?  & $3.94_{-0.28}^{+0.29}$  & $5.43_{-0.60}^{+0.69}$   & 1.02 (514) & $0.89_{-0.22}^{+0.22}$ \\
\hline
\end{tabular}
\tablefoot{Comparison of spectral properties across six source categories fitted by XSPEC model of \texttt{tbabs}$\times$\texttt{apec}. For LPV and LPV candidates, the EW of 6.7 keV line are quantified.}
\end{table}

\subsection{Nature of the X-ray luminous LPVs}

Based on their X-ray spectral properties and optical photometry, we suggest that these X-ray luminous LPVs show characteristics that primarily align with those of $\delta$-type symbiotic stars.
As defined in \citet{Luna2013}, $\delta$-type SySts are characterized as highly absorbed, hard X-ray-emitting sources (E>2.4 keV) that frequently exhibit iron emission lines in the $6.0\text{--}7.0$ keV range. In these systems, the X-ray emission is thought to originate from the boundary layer between an accretion disk and the WD surface.
While the presence of an accretion disk is supported by the detection of iron line complexes \citep{Luna2013, Eze2014}, the exact mechanism powering the mass transfer onto the WD remains an open question \citep{Toala2024}. These systems may be fueled by a standard Bondi-Hoyle wind accretion process, a Roche-lobe overflow (RLOF) similar to those in non-magnetic CVs, or a wind Roche-lobe overflow channel \citep{Bondi1944, Podsiadlowski2007}. 
The specific accretion mode is a primary factor in determining the density of the circumstellar environment and the resulting X-ray signatures. 

In typical SySts \citep{Luna2013}, intense UV radiation from the WD ionizes the dense circumstellar wind of the giant, producing a strong H$\alpha$ excess and a fluorescent $6.4$ keV iron line via reflection of hard X-rays from neutral material. 
However, our sample exhibits a lack of $6.4 \text{ keV}$ Fe \textsc{K}$\alpha$ emission in the stacked X-ray spectra. 
To further investigate the SySt nature of our sample, we cross-matched it with the VPHAS+ survey \citep{Drew2014}. Thanks to the precise astrometry from \gaia, we successfully identified counterparts for all 107 sources in the VPHAS+ catalogs. Notably, only 16 out of 107 sources exhibit a potential H$\alpha$ excess with $(r - \rm H\alpha) > 0.8$, as detailed in Table~\ref{tab:specsrc}. 
Physically, this dual deficiency, the lack of both H$\alpha$ and the 6.4 keV line, suggests that these objects could be accretion-only SySts operating in a relatively "clean" environment with low mass-loss rates from the donor \citep{Mukai2016}. Recent modeling by \cite{Toala2024} demonstrates that a simple disk-like model can account for  the diverse X-ray properties of SySts in which $\delta$-type emission is naturally produced by an accretion disk viewed at near edge-on angles. Crucially, the $6.4\text{ keV}$ fluorescent line becomes negligible under conditions of moderate plasma temperatures ($\rm kT < 10\text{ keV}$) and absorption ($N_{\rm H} < 5\times 10^{22}\text{ cm}^{-2}$) \citep{Toala2024}, consistent with our observations. 
These characteristics suggest that our sources are consistent with the "optically inconspicuous" SySts that emit in the hard X-ray regime, a large hidden population of SySts powered by accretion alone, lacking the WD surface nuclear burning found in more luminous counterparts \citep{Munari2019}.

To summarize, the hard X-ray spectra and optical properties of these luminous late-type giants are highly compatible with accretion-powered processes onto a compact object. We tentatively classify these sources as SySt candidates based on the available data, while noting that further multi-wavelength evidence, such as UV spectroscopy, is required to definitively constrain the nature of the accretors and rule out alternative interpretations.

%\todo[inline,color=red]{more arguments needed?}
\section{Contribution to the GRXE iron line emission}
The X-ray spectra of these SySt candidates, characterized notably by prominent Fe~\textsc{xxv} emission lines, have significant implications for our understanding of the source populations contributing to the GRXE.
Hard X-ray emission from symbiotic stars, dominated by optically thin thermal plasma with $\rm kT \sim 5-10~keV$ and strong Fe~\textsc{xxv} lines, resembles the composite spectra observed from the GRXE \citep[e.g.,][]{Revnivtsev2009, Yuasa2012}.
Although symbiotic stars are relatively rare compared to ABs or non-magnetic CVs, their high intrinsic luminosities imply that even a modest number of such systems could make a non-negligible contribution to the total X-ray emissivity of the Galactic bulge and disk.

\subsection{Construction of the X-ray luminosity function}
\label{sec:xlf_method}

To assess the total contribution of SySts to the GRXE, we constructed their differential X-ray luminosity function (XLF) based on 107 SySt candidates identified in our sample.
We utilized an implementation of the classical $V_{\max}$ formalism \citep{Schmidt1968}, taking into account both the X-ray sensitivity of our survey and the Gaia optical detection limits.

\begin{figure}[h!]
  \centering
  \includegraphics[width=1.0\hsize]{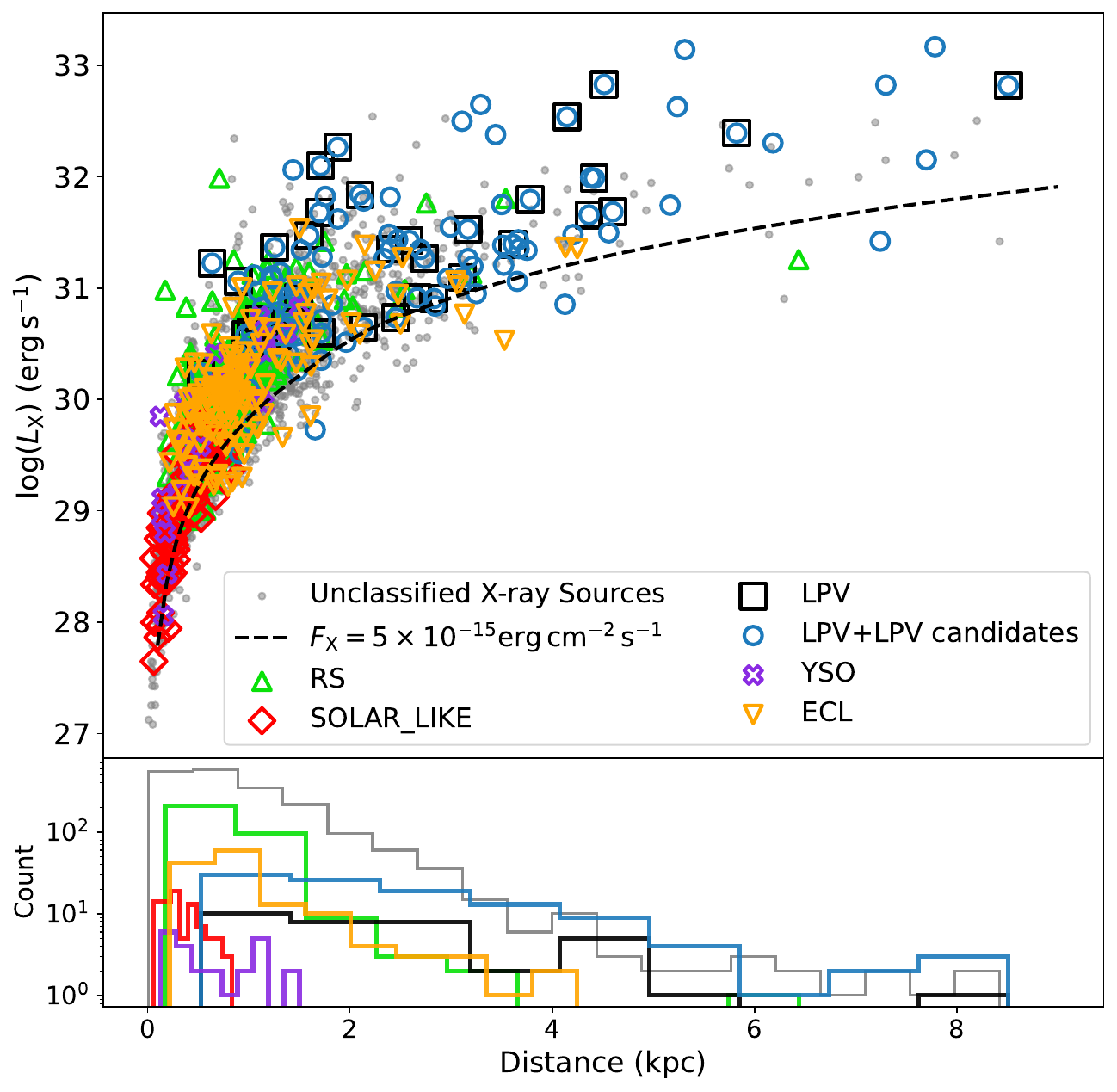}
  \caption{
    Upper panel: X-ray luminosity versus distance. Data points are categorized using the same labels and color scheme defined in Fig.~\ref{fig:CMD_gaiaclass}. The dashed line represent the flux limits, indicating the minimum detectable luminosity at a given distance after accounting for extinction. Lower panel: Distance histograms for the various source class.}
  \label{fig:DistLx}
\end{figure}

For each source, the absorbed X-ray luminosity was computed as $L_{\mathrm{X}} = 4\pi\,d^2\,F_{\mathrm{X}}$,
where $F_{\mathrm{X}}$ is the observed flux in the $0.2$--$12.0$\,keV band and $d$ is the distance derived from \textit{Gaia}~DR3. 
We adopted a survey-wide flux limit of $F_{\mathrm{lim}} = 5 \times 10^{-15}\,\mathrm{erg\,cm^{-2}\,s^{-1}}$. The distribution of their distances is shown in Fig.~\ref{fig:DistLx}, as compared to other classes of sources, showing a greater extension to larger distances. 
While this $F_{\mathrm{lim}}$ does not guarantee 100\% completeness, it provides a conservative space density estimate since a more stringent limit would decrease $V_{\mathrm{max}}$ and consequently yield a higher luminosity function.

The extinction-corrected absolute G-band magnitude was then computed as
\begin{equation}
M_G = G - 5\log_{10}\!\left(\frac{d}{10\,\mathrm{pc}}\right) - A_G.
\end{equation}
For each object we determined the maximum distance at which it remains detectable, accounting for both X-ray and optical constraints. A rigorous upper limit of 20 kpc, corresponding to the expected spatial extent of Galactic sources, was applied in cases where the estimated distance exceeded this threshold (with only one source exceeding the limit in both bands). The X-ray detection limit $F_{\mathrm{X,lim}}$ defines the maximum distance
\begin{equation}
d_{\mathrm{max,X}} = \sqrt{\frac{L_{\mathrm{X}}}{4\pi F_{\mathrm{X,lim}}}},
\end{equation}
while the optical limit $G_{\mathrm{lim}}$ imposes another constraint through
\begin{equation}
G_{\mathrm{lim}} = M_G + 5\log_{10}\!\left(\frac{d_{\mathrm{max,G}}}{10\,\mathrm{pc}}\right) + A_G(d_{\mathrm{max,G}}),
\end{equation}
where the extinction $A_G(d)$ is evaluated along the line of sight. 
The latter equation was solved numerically for each source using a root-finding method, and the effective detection distance was taken as
\begin{equation}
d_{\mathrm{max}} = \min(d_{\mathrm{max,X}},\,d_{\mathrm{max,G}}).
\end{equation}

\begin{figure}[h!]
  \centering
  \includegraphics[width=1.0\hsize]{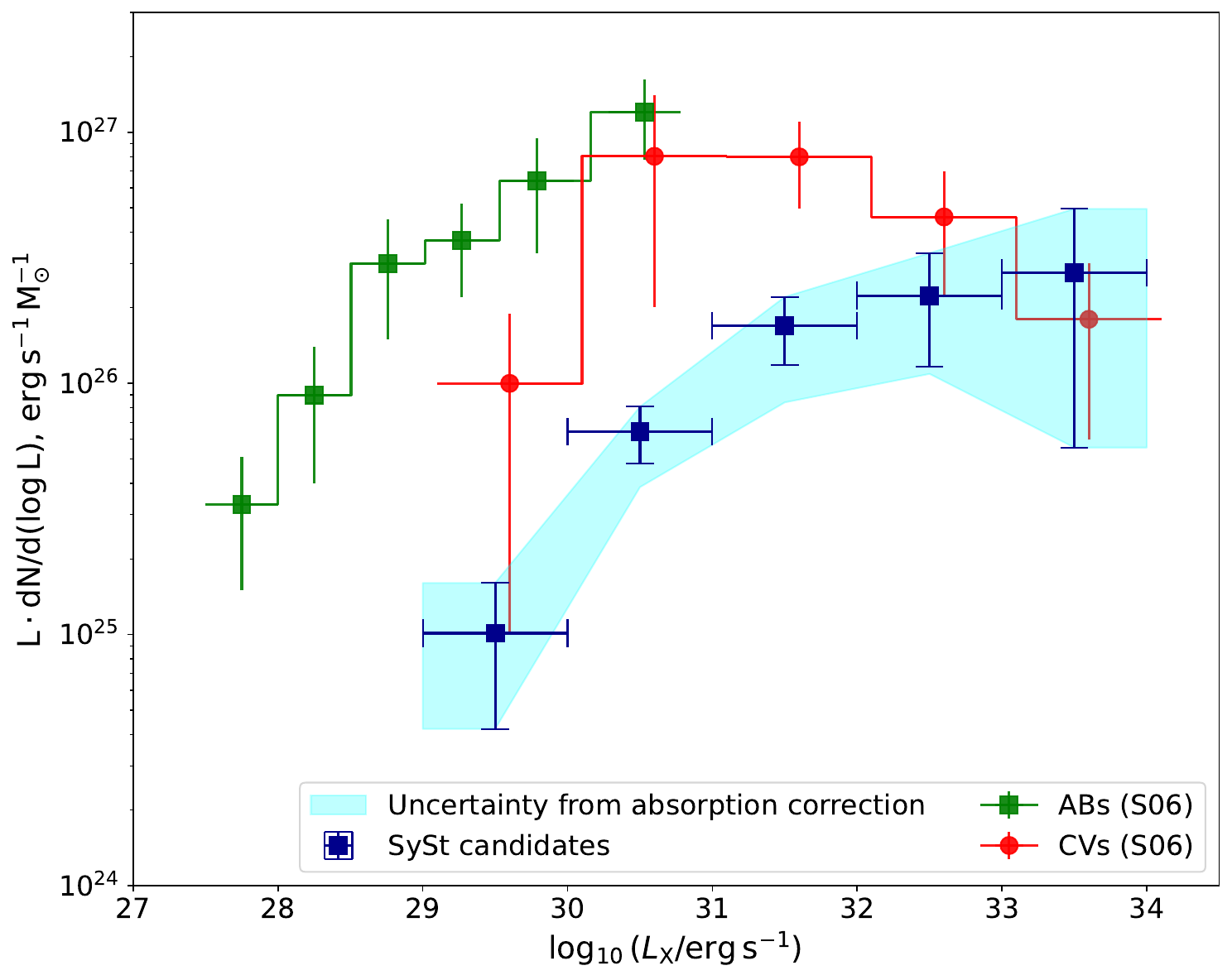}
  \caption{
    Differential XLFs of the SySt candidates identified in this work, normalized to the local stellar mass density. The y-axis represents the luminosity-weighted distribution (contribution per dex in $\log L_{\mathrm{X}}$). For comparison, the XLFs of ABs and CVs from \citet{Sazonov2006} are shown in green and red. The cyan-shaded region represents the systematic uncertainty arising from the choice of extinction model (see text for details).}
  \label{fig:XLF}
\end{figure}

The determination of $d_{\mathrm{max}}$ is sensitive to the choice of extinction modeling. In our primary analysis, we adopted a hybrid approach: for distances up to $3\,\mathrm{kpc}$, we used the 3D dust map of \citet{Lallement2019} to determine the color excess $E(B-V)$ as a function of position (following the procedure in Sect.~\ref{sec:HRX}). For sources located beyond $3\,\mathrm{kpc}$, we transitioned to the model of \citet{Marshall2006}. Since the \citet{Marshall2006} model generally predicts higher extinction than the \citet{Lallement2019} map, a discontinuity may occur at the $3\,\mathrm{kpc}$ boundary. 
To assess the impact of this systematic uncertainty on our results, we calculated an alternative set of values by exclusively employing the \citet{Marshall2006} extinction map across the entire distance range. This approach generally yields a higher X-ray luminosity contribution, as the increased absorption leads to smaller $d_{\mathrm{max}}$ values and, consequently, higher $1/V_{\mathrm{max}}$ weights. These alternative results are shown in Fig.~\ref{fig:XLF} and Tab.~\ref{tab:specsrc}.

The survey covers a solid angle $\Omega \approx 34$ deg$^2$ within the Galactic plane region
($-10^\circ < l < 7^\circ$, $|b| < 1^\circ$). We here adopt a Galactic disk source density profile of,
\begin{equation}
\rho(z,R) \propto \exp \left[-\left(\frac{R_{\mathrm{m}}}{R}\right)^3-\frac{R}{R_{\text {scale }}}-\frac{z}{z_{\text {scale }}}\right],
\end{equation}
where $z = R \sin b$ represents the vertical distance from the plane and $R$ is the Galactocentric radius.
We adopt $R_m$ = 3 kpc, $R_{\rm scale}$ = 3 kpc,  $z_{\rm scale}$ = 200 pc \citep{Binney1997}. 
To facilitate a direct comparison with the results of \citet{Sazonov2006}, we adopt a local stellar mass density ($R = 8.1$ kpc) of $\rho_\ast = 0.04 \, M_{\odot} \, \text{pc}^{-3}$ as the normalization for our model.

For each source the maximum accessible volume is therefore calculated as a
density-weighted volume,
\begin{equation}V_{\mathrm{max},i} = \Omega \int_{0}^{d_{\mathrm{max},i}} \rho(z,R)  R^2 \mathrm{d}R , 
\end{equation}
The differential luminosity function was then estimated using the classical
$1/V_{\mathrm{max}}$ estimator:
\begin{equation}
\frac{dN}{d\log L_{\mathrm{X}}} =
\sum_i \frac{1}{V_{\mathrm{max},i}},
\end{equation}
where the summation is performed within logarithmic luminosity bins of width
$\Delta \log L_{\mathrm{X}} = 1.0$. The statistical uncertainty in each bin is
computed as

\begin{equation}
\sigma =
\sqrt{\sum_i
\left(\frac{1}{V_{\mathrm{max},i}}\right)^2 }.
\end{equation}

Fig.~\ref{fig:XLF} presents the differential XLF of our SySt candidates converted to the 2--10~keV band, alongside the reference XLFs for ABs and CVs from \citet{Sazonov2006}.
Following its methodology, the XLF is expressed as the number of sources per unit solar mass. The cyan-shaded region illustrates the systematic uncertainty associated with the interstellar absorption correction. As discussed before, this uncertainty band encompasses the results obtained from our hybrid extinction approach and a more conservative estimate derived solely from the \citet{Marshall2006} dust map. The latter generally predicts heavier extinction, leading to a higher reconstructed XLF normalization.

\subsection{Cumulative emissivities for SySt candidates}
Using the differential XLF derived in Sect.~\ref{sec:xlf_method}, we computed the cumulative X-ray emissivity of the SySt candidates in the 2--10~keV band. The emissivity per unit stellar mass is defined as
\begin{equation}
\varepsilon = \frac{L_{\mathrm{X}}\cdot dN}{d\log L_{\mathrm{X}}}=\sum_i \frac{L_{\mathrm{X},i}}{V_{\mathrm{max},i}},
\end{equation}
in units of erg\,s$^{-1}$\,$M_{\rm \odot}^{-1}$. 
For our sample, we obtain a cumulative emissivity in the 2-10 keV band of $7.4_{-2.5}^{+2.5} \times10^{26} \rm~erg~s^{-1}~M_{\odot}^{-1}$, or $6.4_{-2.0}^{+2.0} \times10^{26} \rm~erg~s^{-1}~M_{\odot}^{-1}$ in the 3-20 keV band, converted by the best-fit spectral model.
For comparison, the cumulative emissivity of the GRXE in the 3--20~keV band is
 $3.5_{-0.5}^{+0.5} \times10^{27} \rm~erg~s^{-1}~M_{\odot}^{-1}$ \citep{Revnivtsev2006}, implying that SySts contribute $\sim 18\%$ of the total ridge emission per unit stellar mass in this energy band.

More specifically, in the narrow $6.5$--$7.1\text{ keV}$ band encompassing the Fe \textsc{xxv} line, approximately 84\% of the GRXE flux has been resolved into point sources \citep{Revnivtsev2009}. In the $2$--$10\text{ keV}$ range, the emissivity of SySts corresponds to roughly 24\% of the combined emissivity of ABs and CVs ($2.0 \pm 0.8 \times 10^{27}$ and $1.1 \pm 0.3 \times 10^{27}\text{ erg s}^{-1}\text{ M}_{\odot}^{-1}$, respectively, summarized in Tab.~\ref{tab:source_emissivity}). Assuming the resolved point-source population is dominated by these three classes, we estimate that SySts contribute $\sim 20\%$ to the total $2$--$10\text{ keV}$ band. Most notably, because the $6.7\text{ keV}$ EW of SySts is approximately twice that of CVs, their contribution to the narrow $6.5$--$7.1\text{ keV}$ band rises to $\sim 40\%$.

However, systematic uncertainties remain. We derived the XLF for the 107-source sample using the density-weighted $1/V_{\max}$ method, normalized to a local stellar mass density of $\rho_\ast = 0.04 \, M_{\odot} \text{ pc}^{-3}$ for consistency with \citet{Sazonov2006}. Unlike the study by \citet{Sazonov2006}, which relied primarily on a local sample ($d < 1$~kpc), our sample extends to greater distances. Consequently, our density model incorporates both the vertical exponential decay and a radial exponential component for the disk. A more sophisticated stellar mass model (accounting for Galactic bulge, bar and center) is beyond the scope of this work as we are comparing with local, X-ray faint source samples. Furthermore, the homogeneity of our sample remains unconfirmed. While these X-ray-luminous late-type giants are suggested to be accretion-powered SySts, the lack of UV color data precludes a definitive confirmation of the nature of the accretor.

Nevertheless, we did a consistency check by limiting our sample and $d_{\rm max}$ to 1.5 kpc. As shown in Fig.~\ref{fig:DistLx} and Table~\ref{tab:specsrc}, our survey is nearly complete for this type of sources within this volume, as the maximum detectable distances for both optical and X-ray observations generally exceed $1.5$~kpc.
This subsample of 38 sources yields an emissivity of $6.0^{+2.1}_{-2.1} \times 10^{26} \text{ erg s}^{-1} M_{\odot}^{-1}$, which is consistent with the value obtained from the full sample within the uncertainties.

\subsection{Reconciling the GRXE iron line discrepancy}

\begin{figure}[h]
  \centering
  \includegraphics[width=1.0\hsize]{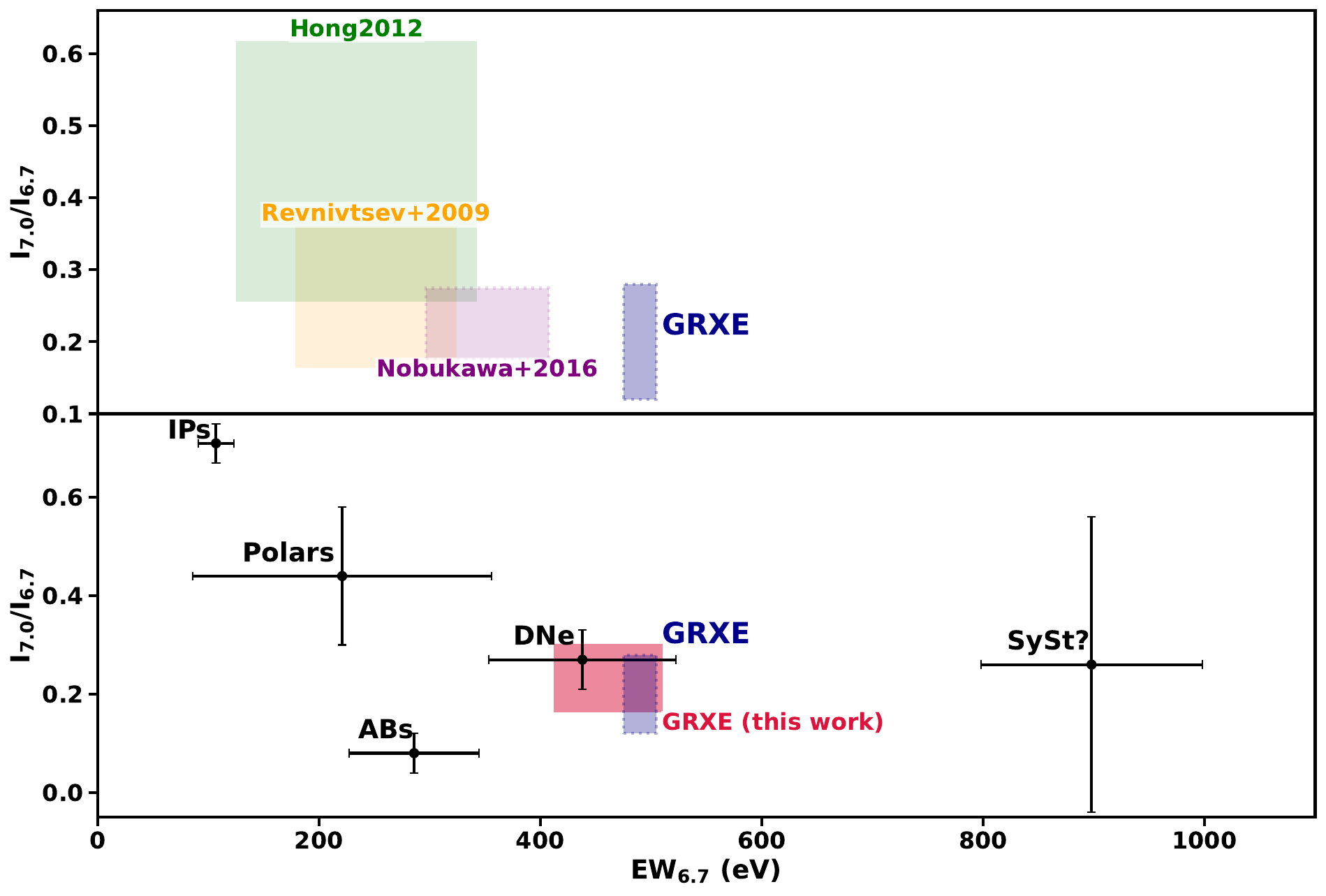}
  \caption{Line ratio $I_{7.0}/I_{6.7}$ vs. EW of the $6.7$~keV line. Upper panel: Comparison of previous GRXE models \citep{Revnivtsev2009, Hong2012, Nobukawa2016} with observed GRXE (blue box). Lower panel: Individual source class measurements from \citet{Xu2016}. The red region shows our updated model (based on \citealt{Nobukawa2016}), which successfully reproduces the GRXE properties by incorporating a 20\% contribution from SySts. }
  \label{fig:EW67}
\end{figure}

Previous studies have suggested that the GRXE is dominated by a combination of non-magnetic CVs like DNe (non-mCVs), magnetic CVs (mCVs), and ABs, though their relative contributions remain a subject of debate. \citet{Revnivtsev2009} attributed 60\% of the GRXE to ABs and 40\% to mCVs. Conversely, \citet{Hong2012} argued for a higher mCV contribution ($\sim$80\%) with less than 20\% from ABs. More recently, \citet{Nobukawa2016} proposed a mixture of 51\% non-mCVs, 10\% mCVs, and 39\% ABs. To test these models, we adopted the measured values for the $\text{EW}_{6.7}$ and the line ratio $I_{7.0}/I_{6.7}$ from \citet{Xu2016} and calculated the resulting parameter space for each prediction. As illustrated in the upper panel of Fig.~\ref{fig:EW67}, a distinct gap persists between these theoretical predictions and the observed GRXE line EW. However, by incorporating a 21\% contribution from our SySt candidates into the population synthesis, specifically adjusting the \citet{Nobukawa2016} model to 20\% SySts, 41\% non-mCVs, 8\% mCVs, and 31\% ABs, we derive an estimated GRXE EW of $467 \pm 50$~eV. This result is in good agreement with the observed value of $490 \pm 15$~eV.

\section{Discussion}
\subsection{The estimated and observed space density of accreting-WD systems}

The space density of CVs remains a subject of active debate due to a persistent, order-of-magnitude discrepancy between theoretical predictions and observational constraints. Population synthesis models consistently predict an overabundance of CVs relative to observations; for instance, models utilizing a classical canonical angular momentum loss (CAML) prescription predict densities as high as $\rho \approx 9 \times 10^{-5}$ pc$^{-3}$, though this can be reconciled to $\rho \approx 2 \times 10^{-5}$ pc$^{-3}$ by adopting empirical CAML prescriptions \citep{Belloni2018}. 

In contrast, observational surveys yield significantly lower values. X-ray flux-limited studies estimate a space density for non-magnetic CVs of $4^{+6}_{-2} \times 10^{-6}$ pc$^{-3}$ \citep{Pretorius2012} and $8^{+4}_{-2} \times 10^{-7}$ pc$^{-3}$ for magnetic CVs (mCVs) \citep{Pretorius2013}. Furthermore, \citet{Revnivtsev2008} derived an even lower value of $1.5^{+0.6}_{-0.6} \times 10^{-7}$ pc$^{-3}$. Recent volume-limited studies of the local 150 pc sample have begun to converge on values between $3.7 ^{+0.7}_{-0.7} \times 10^{-6}$ pc$^{-3}$ \citep{Rodriguez2025} and $4.8^{+0.6}_{-0.8} \times 10^{-6}$ pc$^{-3}$ \citep{Pala2020}.

The space density of SySts is similarly contested, primarily due to uncertainties regarding SySt lifetimes and the binary evolutionary pathways that lead to their formation. Early theoretical estimates for the total Galactic population range from  $4\times10^3$ \citep{Kenyon1986} to $4\times10^5$ \citep{Magrini2003}, which translates to a local space density of $1.0 \times 10^{-8}$ to $1.0 \times 10^{-6}$ pc$^{-3}$ (assuming a Galactic disk radius of 10 kpc).
More recently, \citet{Laversveiler2025} proposed an upper limit of $5.3 \times 10^4$ population from theoretical model, yielding a local density of approximately $1.3 \times 10^{-7}$ pc$^{-3}$. This aligns well with the observational volume density of $\sim (1\text{--}3) \times 10^{-7}$ pc$^{-3}$ derived from a local 500 pc sample using \textit{Gaia}, \textit{GALEX}, and \textit{XMM-Newton} \citep{Xu2024}.
Consequently, despite significant uncertainties, current evidence points toward a consensus value on the order of $10^{-7}$ pc$^{-3}$.

\subsection{Implications of the high number density of SySt candidates}
\label{subsec:discuss}
From the XLF constructed in this work using $V_{\rm max}$ corrections, we derive a total number density of $2.3^{+0.9}_{-0.9} \times 10^{-6}$ pc$^{-3}$ (corresponding to a mass-normalized density of $\rho_M\approx5.8^{+1.9}_{-1.9} \times 10^{-5} \rm ~M_{\odot}^{-1}$), assuming a Galactic scale height of 200 pc. This result is remarkably high, as it is comparable to the observed space density of the entire CV population. This suggests that our current selection, while effective at identifying potential iron line emitters, may encompass a broader and more diverse population than traditionally classified SySts.

\begin{figure}[h!]
  \centering
  \includegraphics[width=1.0\hsize]{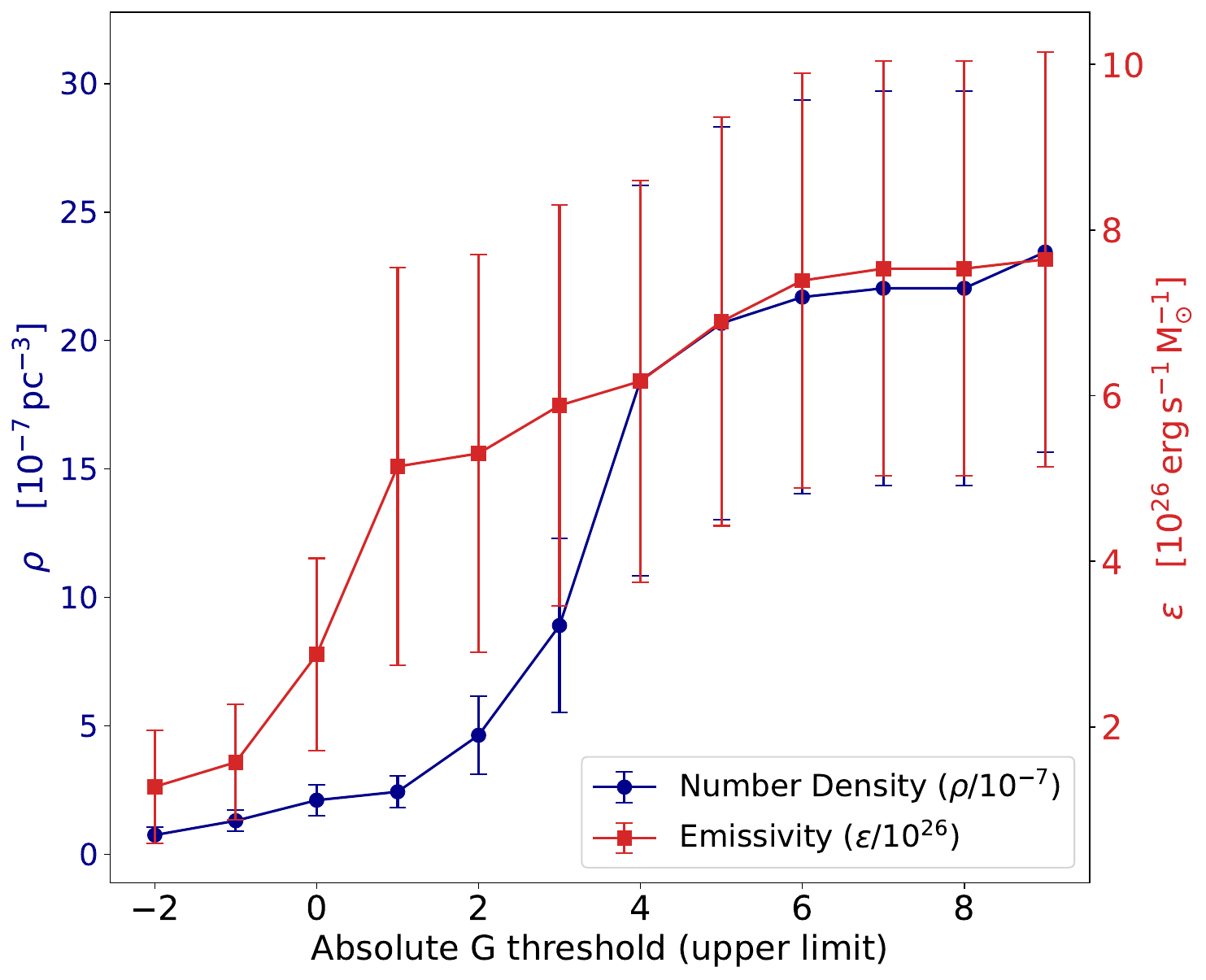}
  \caption{The estimated number density ($\rho$) and emissivity ($\varepsilon$) of subsample derived by putting an upper limit on absolute G magnitude (thereby producing a relatively more luminous sample). }
  \label{fig:rho}
\end{figure}

To investigate the impact of our source selection criteria, we examined how the $M_G$ cuts influence the resulting space density and emissivity (Fig.~\ref{fig:rho}).
Crucially, while total number density is highly sensitive to the inclusion of less luminous sources, the overall contribution to the Galactic Fe~{\sc xxv} emission is primarily driven by the most luminous candidates, whose SySt nature is more robust. For instance, a subsample with $M_G < 1.0$  represents only 10\% of the total number density ($\rho \approx 2.4 \times 10^{-7}$ pc$^{-3}$) yet accounts for $\sim 70\%$ of the total emissivity ($\varepsilon \approx 5.2 \times 10^{26}$ erg s$^{-1} \rm M_{\odot}^{-1}$).

This decoupling of number density from total emissivity indicates that our sample comprises a more heterogeneous population than typical SySts.
While their positions on the \textit{Gaia} CMD, far away from the low-mass main sequence, preclude them from being traditional CVs, they likely represent a more heterogeneous class of interacting binaries.
Recent X-ray studies of AGB stars in the eRASS1 catalog (so-called X-AGBs) reveal a population that may overlap with our low-luminosity SySt candidates. A comparison of X-ray and far-UV luminosities between X-AGBs and X-ray-emitting symbiotic stars (X-SySts) shows a shared luminosity range of $10^{29.5} < L_X < 10^{33.0}$ erg s$^{-1}$ \citep{Guerrero2024}. 
While these X-SySts typically exhibit higher average luminosities ($\sim 10^{32}$ erg s$^{-1}$) attributed to the boundary layer of a white dwarf accretion disk, the lower-luminosity X-AGBs ($\sim 5 \times 10^{30}$ erg s$^{-1}$) may involve accretion onto main-sequence or subgiant F-K companions \citep{Sahai2015}. 
Consequently, the space density of SySts would be notably lower than our total integrated estimate and more aligned with the literature. 
Nevertheless, we emphasize that despite the numerical abundance of these less-luminous sources, their collective contribution to the GRXE remains secondary. The total X-ray emission from this late-type giants population continues to be dominated by the more luminous SySt candidates, reinforcing our conclusion regarding their significant role in the Galactic X-ray background.

Alternatively, those low-luminosity ``X-AGB'' stars may represent a substantial population of SySts that has been largely underestimated in previous studies.
According to \citet{Mukai2016} and their subsequent studies, there should be more "accretion-only" SySts (with luminosities around $\rm 10^{32}~erg~s^{-1}$ or below).
Our spectral analysis offers a clearer perspective on this possibility than previous eRASS1 data, as the detected high-temperature components ($kT \sim 4\text{--}5$~keV) and Fe line features are consistent with accretion onto a white dwarf boundary layer. This interpretation aligns with the findings of \citet{Ortiz2021} for X-AGB stars in the 4XMM-DR9 catalog, who noted that plasma temperatures exceeding $10^7$~K are likely indicative of a symbiotic nature involving a compact companion.
If this is the case, the unexpectedly high space density of our SySt candidates compared to traditional models can be attributed to two main factors. First, unlike CVs, which require precise Roche-lobe overflow, SySt systems can initiate X-ray emission via wind accretion across a much broader range of orbital periods. While CVs (a main-sequence star and a WD) must satisfy narrow constraints to avoid either premature merging or accretion failure, the massive winds from giants allow WDs to capture material even in wide orbits. This "wind-capture" flexibility likely accounts for a larger active population than RLOF-limited models suggest. Second, current theory on WDs accreting from giants is still quite underdeveloped. One example is that the observed bolometric luminosities (UV + X-ray) are frequently higher than predicted, forcing models to assume enhanced wind-loss or higher accretion rates \citep{Meng2016, Yu2022}. These uncertainties suggest that existing theoretical predictions for SySt space densities may be underestimated.

In summary, we suggest this 107-source SySt candidates sample may represent a mixture of ``accretion-only'' SySts as defined by \citet{Mukai2016}, and the ``X-AGB'' stars as also identified in the \textit{eRASS1} catalog \citep{Guerrero2024}. 
If the latter involve accretion onto main-sequence companions rather than WDs, the SySt space density would be lower and more aligned with historical values.
However, if these sources are indeed symbiotic systems, our results provide a valuable empirical constraint suggesting that current theoretical models may significantly underestimate the population of SySts.
Crucially, regardless of the exact nature of the fainter members, the total Galactic Fe~{\sc xxv} emissivity of these X-ray luminous late-giants is fundamentally driven by the most luminous SySt candidates.
Future multi-wavelength follow-ups, particularly UV spectroscopy, will be essential to definitively distinguish between these subclasses and to refine the specific contribution of SySts to the Galactic X-ray background.

\begin{table}[t] 
\centering 
\caption{Emissivities of different classes of sources.}
\label{tab:source_emissivity}
\setlength{\tabcolsep}{1pt}
\renewcommand{\arraystretch}{1.4} 
\begin{tabular}{llcc}
\toprule
\toprule
Class & EW 6.7 & \multicolumn{2}{c}{Total emissivity ($10^{27}$ erg s$^{-1}M_{\odot}^{-1}$)} \\
\cmidrule(lr){3-4}
      & (eV)   & 2--10 keV & 3--20 keV \\
\midrule
ABs\tablefootmark{a} & $286 \pm 59$ & $2.0 \pm 0.8$ & $2.9 \pm 1.3$ \\
\addlinespace[0.5em]
CVs     &                       & $1.1 \pm 0.3$ & $2.4 \pm 0.6$ \\
        & IPs: $107 \pm 16$     & \dots         & \dots         \\
        & Polars: $221 \pm 135$ & \dots         & \dots         \\
        & DNe: $438 \pm 85$     & \dots         & \dots         \\
\addlinespace[0.5em]
ABs + CVs & \dots               & $3.1 \pm 0.8$ & $5.3 \pm 1.5$ \\
\textbf{SySt?}  & $\mathbf{890 \pm 100}$ & $\mathbf{0.7 \pm 0.3}$ & $\mathbf{0.6 \pm 0.2}$ \\
\midrule
GRXE\tablefootmark{b} & $490 \pm 15$ & \dots & $3.5 \pm 0.5$ \\
\bottomrule
\end{tabular}

\tablefoot{
\tablefoottext{a}{\citet{Sazonov2006}, based on an RXTE sample within 1 kpc.}
\tablefoottext{b}{\citet{Revnivtsev2006}.}
The emissivity values for SySt candidates derived in this work are highlighted in bold. The values of EW 6.7 for individual source classes and GRXE are adopted from \citep{Xu2016}.}
\end{table}

%%%%%%%%%%%%%%%%%%%%%%%%%%%%%%%%%%%%%%%%%%%%%%%%%%%%%%%%%%%%%%
\section{Conclusions}

Our results provide a new perspective on the origin of the Galactic ridge X-ray emission. While it has been widely accepted that the GRXE is primarily dominated by CVs along with contributions from ABs, the specific origin of its high-energy spectral features remained a subject of debate. By identifying 107 X-ray luminous sources associated with late-type giant stars, primarily suggested to be symbiotic stars, we have bridged a significant gap in accounting for the Galactic ridge $6.7 \text{ keV}$ iron line flux.
Our analysis indicates that this population contributes approximately $40\%$ of the iron line emission of the Galactic ridge. This finding suggests that X-ray luminous late-type giants are one of the main contributors to the high-energy features of the GRXE rather than mere outliers. 
These results fit into a wider context where the GRXE is increasingly understood as the integrated emission of various discrete stellar populations rather than a truly diffuse hot interstellar medium. 
Future studies using next-generation X-ray observatories with higher spectral resolution such as \textit{NewAthena} \citep{NewAthena}, will be essential to further refine these contributions. Furthermore, multi-wavelength follow-ups of our candidate sources will help constrain the total space density of these X-ray luminous giants, ultimately completing the census of the Milky Way high-energy background.

%Future studies should assess the integrated contribution of these evolved binaries to the Galactic X-ray background.
%%%%%%%%%%%%%%%%%%%%%%%%%%%%%%%%%%%%%%%%%%%%%%%%%%%%%%%%%%%%%%
\begin{acknowledgements}
T.B. and G.P acknowledge financial support from Bando per il Finanziamento della Ricerca Fondamentale 2022 dell’Istituto Nazionale di Astrofisica (INAF): GO Large program, and from the Framework per l'Attrazione e il Rafforzamento delle Eccellenze (FARE) per la ricerca in Italia (R20L5S39T9).
G.P. also acknowledges financial support from the European Research Council (ERC) under the European Union's Horizon 2020 research and innovation program HotMilk (grant agreement No. 865637).
MRM acknowledges support from NASA ADAP grant 80NSSC24K0639.
TMD and MAPT acknowledge support by the Spanish \textit{Agencia estatal de investigaci\'on} via PID2021-124879NB-I00 and PID2024-161863NB-I00.
\end{acknowledgements}

% Please note that we have included the references below in
% order to compile the document, but we ask you to:
\bibliographystyle{aa} % style aa.bst
\bibliography{refs.bib} % your references refs.bib

@ARTICLE{Worrall1983,
       author = {{Worrall}, D.~M. and {Marshall}, F.~E.},
        title = "{Stellar contributions to the hard X-ray galactic ridge.}",
      journal = {\apj},
         year = 1983,
        month = apr,
       volume = {267},
        pages = {691-697},
          doi = {10.1086/160906},
       adsurl = {https://ui.adsabs.harvard.edu/abs/1983ApJ...267..691W}
}

@ARTICLE{Koyama1986,
       author = {{Koyama}, K. and {Makishima}, K. and {Tanaka}, Y. and {Tsunemi}, H.},
        title = "{Thermal X-Ray Emission with Intense 6.7-keV Iron Line from the Galactic Ridge}",
      journal = {\pasj},
         year = 1986,
        month = mar,
       volume = {38},
       number = {1},
        pages = {121-131},
          doi = {10.1093/pasj/38.1.121},
       adsurl = {https://ui.adsabs.harvard.edu/abs/1986PASJ...38..121K}
}

@ARTICLE{struder2001,
    author = {{Str{\"u}der}, L. and {Briel}, U. and {Dennerl}, K. and {Hartmann}, R. and 
        {Kendziorra}, E. and {Meidinger}, N. and {Pfeffermann}, E. and 
        {Reppin}, C. and {Aschenbach}, B. and {Bornemann}, W. and {Br{\"a}uninger}, H. and 
        {Burkert}, W. and {Elender}, M. and {Freyberg}, M. and {Haberl}, F. and 
        {Hartner}, G. and {Heuschmann}, F. and {Hippmann}, H. and {Kastelic}, E. and 
        {Kemmer}, S. and {Kettenring}, G. and {Kink}, W. and {Krause}, N. and 
        {M{\"u}ller}, S. and {Oppitz}, A. and {Pietsch}, W. and {Popp}, M. and 
        {Predehl}, P. and {Read}, A. and {Stephan}, K.~H. and {St{\"o}tter}, D. and 
        {Tr{\"u}mper}, J. and {Holl}, P. and {Kemmer}, J. and {Soltau}, H. and 
        {St{\"o}tter}, R. and {Weber}, U. and {Weichert}, U. and {von Zanthier}, C. and 
        {Carathanassis}, D. and {Lutz}, G. and {Richter}, R.~H. and 
        {Solc}, P. and {B{\"o}ttcher}, H. and {Kuster}, M. and {Staubert}, R. and 
        {Abbey}, A. and {Holland}, A. and {Turner}, M. and {Balasini}, M. and 
        {Bignami}, G.~F. and {La Palombara}, N. and {Villa}, G. and 
        {Buttler}, W. and {Gianini}, F. and {Lain{\'e}}, R. and {Lumb}, D. and 
        {Dhez}, P.},
    title = "The European Photon Imaging Camera on XMM-Newton:   The pn-CCD camera",
    journal = {\aap},
    year = 2001,
    month = jan,
    volume = 365,
    pages = {L18--L26},
    url = {http://cdsads.u-strasbg.fr/cgi-bin/nph-bib_query?bibcode=2001A%26A...365L..18S&db_key=AST}
}

@ARTICLE{turner2001,
    author = {{Turner}, M.~J.~L. and {Abbey}, A. and {Arnaud}, M. and {Balasini}, M. and 
        {Barbera}, M. and {Belsole}, E. and {Bennie}, P.~J. and {Bernard}, J.~P. and 
        {Bignami}, G.~F. and {Boer}, M. and {Briel}, U. and {Butler}, I. and 
        {Cara}, C. and {Chabaud}, C. and {Cole}, R. and {Collura}, A. and 
        {Conte}, M. and {Cros}, A. and {Denby}, M. and {Dhez}, P. and 
        {Di Coco}, G. and {Dowson}, J. and {Ferrando}, P. and {Ghizzardi}, S. and 
        {Gianotti}, F. and {Goodall}, C.~V. and {Gretton}, L. and {Griffiths}, R.~G. and 
        {Hainaut}, O. and {Hochedez}, J.~F. and {Holland}, A.~D. and 
        {Jourdain}, E. and {Kendziorra}, E. and {Lagostina}, A. and 
        {Laine}, R. and {La Palombara}, N. and {Lortholary}, M. and 
        {Lumb}, D. and {Marty}, P. and {Molendi}, S. and {Pigot}, C. and 
        {Poindron}, E. and {Pounds}, K.~A. and {Reeves}, J.~N. and 
        {Reppin}, C. and {Rothenflug}, R. and {Salvetat}, P. and {Sauvageot}, J.~L. and 
        {Schmitt}, D. and {Sembay}, S. and {Short}, A.~D.~T. and {Spragg}, J. and 
        {Stephen}, J. and {Str{\"u}der}, L. and {Tiengo}, A. and {Trifoglio}, M. and 
        {Tr{\"u}mper}, J. and {Vercellone}, S. and {Vigroux}, L. and 
        {Villa}, G. and {Ward}, M.~J. and {Whitehead}, S. and {Zonca}, E.
        },
    title = "The European Photon Imaging Camera on XMM-Newton:   The MOS cameras : The MOS cameras",
    journal = {\aap},
    year = 2001,
    month = jan,
    volume = 365,
    pages = {L27--L35},
    url = {http://cdsads.u-strasbg.fr/cgi-bin/nph-bib_query?bibcode=2001A%26A...365L..27T&db_key=AST}
}

@ARTICLE{Lallement2019,
       author = {{Lallement}, R. and {Babusiaux}, C. and {Vergely}, J.~L. and {Katz}, D. and {Arenou}, F. and {Valette}, B. and {Hottier}, C. and {Capitanio}, L.},
        title = "{Gaia-2MASS 3D maps of Galactic interstellar dust within 3 kpc}",
      journal = {\aap},
         year = 2019,
        month = may,
       volume = {625},
          eid = {A135},
        pages = {A135},
          doi = {10.1051/0004-6361/201834695},
archivePrefix = {arXiv},
       eprint = {1902.04116},
 primaryClass = {astro-ph.GA},
       adsurl = {https://ui.adsabs.harvard.edu/abs/2019A&A...625A.135L}
}

@ARTICLE{GaiaDR3,
       author = {{Gaia Collaboration} and {Vallenari}, A. and {Brown}, A.~G.~A. and {Prusti}, T. and {de Bruijne}, J.~H.~J. and {Arenou}, F. and {Babusiaux}, C. and {Biermann}, M. and {Creevey}, O.~L. and {Ducourant}, C. and {Evans}, D.~W. and {Eyer}, L. and {Guerra}, R. and {Hutton}, A. and {Jordi}, C. and {Klioner}, S.~A. and {Lammers}, U.~L. and {Lindegren}, L. and {Luri}, X. and {Mignard}, F. and {Panem}, C. and {Pourbaix}, D. and {Randich}, S. and {Sartoretti}, P. and {Soubiran}, C. and {Tanga}, P. and {Walton}, N.~A. and {Bailer-Jones}, C.~A.~L. and {Bastian}, U. and {Drimmel}, R. and {Jansen}, F. and {Katz}, D. and {Lattanzi}, M.~G. and {van Leeuwen}, F. and {Bakker}, J. and {Cacciari}, C. and {Casta{\~n}eda}, J. and {De Angeli}, F. and {Fabricius}, C. and {Fouesneau}, M. and {Fr{\'e}mat}, Y. and {Galluccio}, L. and {Guerrier}, A. and {Heiter}, U. and {Masana}, E. and {Messineo}, R. and {Mowlavi}, N. and {Nicolas}, C. and {Nienartowicz}, K. and {Pailler}, F. and {Panuzzo}, P. and {Riclet}, F. and {Roux}, W. and {Seabroke}, G.~M. and {Sordo}, R. and {Th{\'e}venin}, F. and {Gracia-Abril}, G. and {Portell}, J. and {Teyssier}, D. and {Altmann}, M. and {Andrae}, R. and {Audard}, M. and {Bellas-Velidis}, I. and {Benson}, K. and {Berthier}, J. and {Blomme}, R. and {Burgess}, P.~W. and {Busonero}, D. and {Busso}, G. and {C{\'a}novas}, H. and {Carry}, B. and {Cellino}, A. and {Cheek}, N. and {Clementini}, G. and {Damerdji}, Y. and {Davidson}, M. and {de Teodoro}, P. and {Nu{\~n}ez Campos}, M. and {Delchambre}, L. and {Dell'Oro}, A. and {Esquej}, P. and {Fern{\'a}ndez-Hern{\'a}ndez}, J. and {Fraile}, E. and {Garabato}, D. and {Garc{\'\i}a-Lario}, P. and {Gosset}, E. and {Haigron}, R. and {Halbwachs}, J. -L. and {Hambly}, N.~C. and {Harrison}, D.~L. and {Hern{\'a}ndez}, J. and {Hestroffer}, D. and {Hodgkin}, S.~T. and {Holl}, B. and {Jan{\ss}en}, K. and {Jevardat de Fombelle}, G. and {Jordan}, S. and {Krone-Martins}, A. and {Lanzafame}, A.~C. and {L{\"o}ffler}, W. and {Marchal}, O. and {Marrese}, P.~M. and {Moitinho}, A. and {Muinonen}, K. and {Osborne}, P. and {Pancino}, E. and {Pauwels}, T. and {Recio-Blanco}, A. and {Reyl{\'e}}, C. and {Riello}, M. and {Rimoldini}, L. and {Roegiers}, T. and {Rybizki}, J. and {Sarro}, L.~M. and {Siopis}, C. and {Smith}, M. and {Sozzetti}, A. and {Utrilla}, E. and {van Leeuwen}, M. and {Abbas}, U. and {{\'A}brah{\'a}m}, P. and {Abreu Aramburu}, A. and {Aerts}, C. and {Aguado}, J.~J. and {Ajaj}, M. and {Aldea-Montero}, F. and {Altavilla}, G. and {{\'A}lvarez}, M.~A. and {Alves}, J. and {Anders}, F. and {Anderson}, R.~I. and {Anglada Varela}, E. and {Antoja}, T. and {Baines}, D. and {Baker}, S.~G. and {Balaguer-N{\'u}{\~n}ez}, L. and {Balbinot}, E. and {Balog}, Z. and {Barache}, C. and {Barbato}, D. and {Barros}, M. and {Barstow}, M.~A. and {Bartolom{\'e}}, S. and {Bassilana}, J. -L. and {Bauchet}, N. and {Becciani}, U. and {Bellazzini}, M. and {Berihuete}, A. and {Bernet}, M. and {Bertone}, S. and {Bianchi}, L. and {Binnenfeld}, A. and {Blanco-Cuaresma}, S. and {Blazere}, A. and {Boch}, T. and {Bombrun}, A. and {Bossini}, D. and {Bouquillon}, S. and {Bragaglia}, A. and {Bramante}, L. and {Breedt}, E. and {Bressan}, A. and {Brouillet}, N. and {Brugaletta}, E. and {Bucciarelli}, B. and {Burlacu}, A. and {Butkevich}, A.~G. and {Buzzi}, R. and {Caffau}, E. and {Cancelliere}, R. and {Cantat-Gaudin}, T. and {Carballo}, R. and {Carlucci}, T. and {Carnerero}, M.~I. and {Carrasco}, J.~M. and {Casamiquela}, L. and {Castellani}, M. and {Castro-Ginard}, A. and {Chaoul}, L. and {Charlot}, P. and {Chemin}, L. and {Chiaramida}, V. and {Chiavassa}, A. and {Chornay}, N. and {Comoretto}, G. and {Contursi}, G. and {Cooper}, W.~J. and {Cornez}, T. and {Cowell}, S. and {Crifo}, F. and {Cropper}, M. and {Crosta}, M. and {Crowley}, C. and {Dafonte}, C. and {Dapergolas}, A. and {David}, M. and {David}, P. and {de Laverny}, P. and {De Luise}, F. and {De March}, R.},
        title = "{Gaia Data Release 3. Summary of the content and survey properties}",
      journal = {\aap},
         year = 2023,
        month = jun,
       volume = {674},
          eid = {A1},
        pages = {A1},
          doi = {10.1051/0004-6361/202243940},
archivePrefix = {arXiv},
       eprint = {2208.00211},
 primaryClass = {astro-ph.GA},
       adsurl = {https://ui.adsabs.harvard.edu/abs/2023A&A...674A...1G}
}

@ARTICLE{Wang2019,
       author = {{Wang}, Shu and {Chen}, Xiaodian},
        title = "{The Optical to Mid-infrared Extinction Law Based on the APOGEE, Gaia DR2, Pan-STARRS1, SDSS, APASS, 2MASS, and WISE Surveys}",
      journal = {\apj},
         year = 2019,
        month = jun,
       volume = {877},
       number = {2},
          eid = {116},
        pages = {116},
          doi = {10.3847/1538-4357/ab1c61},
archivePrefix = {arXiv},
       eprint = {1904.04575},
 primaryClass = {astro-ph.GA},
       adsurl = {https://ui.adsabs.harvard.edu/abs/2019ApJ...877..116W}
}

@ARTICLE{Allen1984,
       author = {{Allen}, D.~A.},
        title = "{A catalogue of symbiotic stars.}",
      journal = {\pasa},
         year = 1984,
        month = jan,
       volume = {5},
       number = {3},
        pages = {369-421},
          doi = {10.1017/S1323358000017215},
       adsurl = {https://ui.adsabs.harvard.edu/abs/1984PASA....5..369A}
}

@ARTICLE{Allen1981,
       author = {{Allen}, D.~A.},
        title = "{X-ray observations of symbiotic stars.}",
      journal = {\mnras},
         year = 1981,
        month = nov,
       volume = {197},
        pages = {739-743},
          doi = {10.1093/mnras/197.3.739},
       adsurl = {https://ui.adsabs.harvard.edu/abs/1981MNRAS.197..739A}
}

@ARTICLE{Belczynski2000,
       author = {{Belczy{\'n}ski}, K. and {Miko{\l}ajewska}, J. and {Munari}, U. and {Ivison}, R.~J. and {Friedjung}, M.},
        title = "{A catalogue of symbiotic stars}",
      journal = {\aaps},
         year = 2000,
        month = nov,
       volume = {146},
        pages = {407-435},
          doi = {10.1051/aas:2000280},
archivePrefix = {arXiv},
       eprint = {astro-ph/0005547},
 primaryClass = {astro-ph},
       adsurl = {https://ui.adsabs.harvard.edu/abs/2000A&AS..146..407B}
}

@ARTICLE{4XMMDR14s,
       author = {{Traulsen}, I. and {Schwope}, A.~D. and {Lamer}, G. and {Ballet}, J. and {Carrera}, F.~J. and {Ceballos}, M.~T. and {Coriat}, M. and {Freyberg}, M.~J. and {Koliopanos}, F. and {Kurpas}, J. and {Michel}, L. and {Motch}, C. and {Page}, M.~J. and {Watson}, M.~G. and {Webb}, N.~A.},
        title = "{The XMM-Newton serendipitous survey. X. The second source catalogue from overlapping XMM-Newton observations and its long-term variable content}",
      journal = {\aap},
         year = 2020,
        month = sep,
       volume = {641},
          eid = {A137},
        pages = {A137},
          doi = {10.1051/0004-6361/202037706},
archivePrefix = {arXiv},
       eprint = {2007.02932},
 primaryClass = {astro-ph.HE},
       adsurl = {https://ui.adsabs.harvard.edu/abs/2020A&A...641A.137T}
}

@ARTICLE{Hong2012,
       author = {{Hong}, JaeSub and {van den Berg}, Maureen and {Grindlay}, Jonathan E. and {Servillat}, Mathieu and {Zhao}, Ping},
        title = "{Discovery of a Significant Magnetic Cataclysmic Variable Population in the Limiting Window}",
      journal = {\apj},
         year = 2012,
        month = feb,
       volume = {746},
       number = {2},
          eid = {165},
        pages = {165},
          doi = {10.1088/0004-637X/746/2/165},
archivePrefix = {arXiv},
       eprint = {1103.2477},
 primaryClass = {astro-ph.HE},
       adsurl = {https://ui.adsabs.harvard.edu/abs/2012ApJ...746..165H}
}

@ARTICLE{Tong2025,
       author = {{Bao}, Tong and {Ponti}, Gabriele and {Haberl}, Frank and {Mondal}, Samaresh and {Morris}, Mark R. and {Mori}, Kaya and {Mandel}, Shifra and {Xu}, Xiao-jie},
        title = "{Unveiling the soft X-ray source population towards the inner Galactic disk with XMM-Newton}",
      journal = {arXiv e-prints},
         year = 2025,
        month = oct,
          eid = {arXiv:2510.23814},
        pages = {arXiv:2510.23814},
          doi = {10.48550/arXiv.2510.23814},
archivePrefix = {arXiv},
       eprint = {2510.23814},
 primaryClass = {astro-ph.HE},
       adsurl = {https://ui.adsabs.harvard.edu/abs/2025arXiv251023814B}
}

@ARTICLE{gaiaclass,
       author = {{Rimoldini}, Lorenzo and {Holl}, Berry and {Gavras}, Panagiotis and {Audard}, Marc and {De Ridder}, Joris and {Mowlavi}, Nami and {Nienartowicz}, Krzysztof and {Jevardat de Fombelle}, Gr{\'e}gory and {Lecoeur-Ta{\"\i}bi}, Isabelle and {Karbevska}, Lea and {Evans}, Dafydd W. and {{\'A}brah{\'a}m}, P{\'e}ter and {Carnerero}, Maria I. and {Clementini}, Gisella and {Distefano}, Elisa and {Garofalo}, Alessia and {Garc{\'\i}a-Lario}, Pedro and {Gomel}, Roy and {Klioner}, Sergei A. and {Kruszy{\'n}ska}, Katarzyna and {Lanzafame}, Alessandro C. and {Lebzelter}, Thomas and {Marton}, G{\'a}bor and {Mazeh}, Tsevi and {Molinaro}, Roberto and {Panahi}, Aviad and {Raiteri}, Claudia M. and {Ripepi}, Vincenzo and {Szabados}, L{\'a}szl{\'o} and {Teyssier}, David and {Trabucchi}, Michele and {Wyrzykowski}, {\L}ukasz and {Zucker}, Shay and {Eyer}, Laurent},
        title = "{Gaia Data Release 3. All-sky classification of 12.4 million variable sources into 25 classes}",
      journal = {\aap},
         year = 2023,
        month = jun,
       volume = {674},
          eid = {A14},
        pages = {A14},
          doi = {10.1051/0004-6361/202245591},
archivePrefix = {arXiv},
       eprint = {2211.17238},
 primaryClass = {astro-ph.GA},
       adsurl = {https://ui.adsabs.harvard.edu/abs/2023A&A...674A..14R}
}

@software{gordon2021,
       author = {{Gordon}, Craig and {Arnaud}, Keith},
        title = "{PyXspec: Python interface to XSPEC spectral-fitting program}",
 howpublished = {Astrophysics Source Code Library, record ascl:2101.014},
         year = 2021,
        month = jan,
          eid = {ascl:2101.014},
       adsurl = {https://ui.adsabs.harvard.edu/abs/2021ascl.soft01014G}
}

@article{Preibisch2005,
  author = {Preibisch, T. and Kim, Y.-C. and Favata, F. and Feigelson, E. D. and Flaccomio, E. and Getman, K. and Micela, G. and Sciortino, S. and Stassun, K. and Stelzer, B. and Zinnecker, H.},
  title = {The Origin of T Tauri X-Ray Emission: New Insights from the Chandra Orion Ultradeep Project},
  journal = {ApJS},
  volume = {160},
  pages = {401--422},
  year = {2005}
}

@article{Getman2005,
  author = {Getman, K. V. and Feigelson, E. D. and Grosso, N. and McCaughrean, M. J. and Micela, G. and Broos, P. and Garmire, G. P. and Townsley, L. and others},
  title = {Chandra Orion Ultradeep Project: Observations and Source Lists},
  journal = {ApJS},
  volume = {160},
  pages = {319--352},
  year = {2005}
}

@article{Gudel2007,
  author = {Güdel, M. and Skinner, S. L. and Briggs, K. R. and Audard, M. and Arzner, K. and Telleschi, A.},
  title = {X-ray emission from young stellar objects},
  journal = {A\&A},
  volume = {468},
  pages = {353--377},
  year = {2007}
}

@article{Luna2013,
  author = {Luna, G. J. M. and Sokoloski, J. L. and Mukai, K. and Nelson, T.},
  title = {The X-ray view of symbiotic stars},
  journal = {A\&A},
  volume = {559},
  pages = {A6},
  year = {2013}
}

@article{Mukai2017,
  author = {Mukai, K.},
  title = {X-ray emission from accreting white dwarfs},
  journal = {PASP},
  volume = {129},
  number = {976},
  pages = {062001},
  year = {2017}
}

@article{Revnivtsev2009,
  author = {Revnivtsev, M. and Sazonov, S. and Churazov, E. and Forman, W. and Vikhlinin, A. and Sunyaev, R.},
  title = {Discrete sources as the origin of the Galactic ridge X-ray emission},
  journal = {Nature},
  volume = {458},
  pages = {1142--1144},
  year = {2009}
}

@article{Yuasa2012,
  author = {Yuasa, T. and Makishima, K. and Nakazawa, K.},
  title = {Revealing the origin of the Galactic Ridge X-ray Emission using Suzaku},
  journal = {ApJ},
  volume = {753},
  pages = {129},
  year = {2012}
}

@article{Xu2016,
  author = {Xu, X. and Wang, Q. D. and Li, X.-D.},
  title = {Galactic Ridge X-ray Emission: Contribution from faint compact binaries},
  journal = {ApJ},
  volume = {818},
  pages = {136},
  year = {2016}
}

@ARTICLE{Schmitt2024,
       author = {{Schmitt}, J.~H.~M.~M. and {H{\"u}nsch}, M. and {Schneider}, P.~C. and {Freund}, S. and {Czesla}, S. and {Robrade}, J. and {Schwope}, A.},
        title = "{``Forbidden'' stars in the eROSITA all-sky survey: X-ray emission from very late-type giants}",
      journal = {\aap},
         year = 2024,
        month = aug,
       volume = {688},
          eid = {A9},
        pages = {A9},
          doi = {10.1051/0004-6361/202449181},
archivePrefix = {arXiv},
       eprint = {2401.17273},
 primaryClass = {astro-ph.SR},
       adsurl = {https://ui.adsabs.harvard.edu/abs/2024A&A...688A...9S}
}

@ARTICLE{Bressan2012,
       author = {{Bressan}, Alessandro and {Marigo}, Paola and {Girardi}, L{\'e}o. and {Salasnich}, Bernardo and {Dal Cero}, Claudia and {Rubele}, Stefano and {Nanni}, Ambra},
        title = "{PARSEC: stellar tracks and isochrones with the PAdova and TRieste Stellar Evolution Code}",
      journal = {\mnras},
         year = 2012,
        month = nov,
       volume = {427},
       number = {1},
        pages = {127-145},
          doi = {10.1111/j.1365-2966.2012.21948.x},
archivePrefix = {arXiv},
       eprint = {1208.4498},
 primaryClass = {astro-ph.SR},
       adsurl = {https://ui.adsabs.harvard.edu/abs/2012MNRAS.427..127B}
}

@ARTICLE{Sazonov2006,
       author = {{Sazonov}, S. and {Revnivtsev}, M. and {Gilfanov}, M. and {Churazov}, E. and {Sunyaev}, R.},
        title = "{X-ray luminosity function of faint point sources in the Milky Way}",
      journal = {\aap},
         year = 2006,
        month = apr,
       volume = {450},
       number = {1},
        pages = {117-128},
          doi = {10.1051/0004-6361:20054297},
archivePrefix = {arXiv},
       eprint = {astro-ph/0510049},
 primaryClass = {astro-ph},
       adsurl = {https://ui.adsabs.harvard.edu/abs/2006A&A...450..117S}
}

@ARTICLE{Schmidt1968,
       author = {{Schmidt}, Maarten},
        title = "{Space Distribution and Luminosity Functions of Quasi-Stellar Radio Sources}",
      journal = {\apj},
         year = 1968,
        month = feb,
       volume = {151},
        pages = {393},
          doi = {10.1086/149446},
       adsurl = {https://ui.adsabs.harvard.edu/abs/1968ApJ...151..393S}
}

@ARTICLE{Revnivtsev2006,
       author = {{Revnivtsev}, M. and {Sazonov}, S. and {Gilfanov}, M. and {Churazov}, E. and {Sunyaev}, R.},
        title = "{Origin of the Galactic ridge X-ray emission}",
      journal = {\aap},
         year = 2006,
        month = jun,
       volume = {452},
       number = {1},
        pages = {169-178},
          doi = {10.1051/0004-6361:20054268},
       adsurl = {https://ui.adsabs.harvard.edu/abs/2006A&A...452..169R}
}

@ARTICLE{Ayres2005,
       author = {{Ayres}, Thomas R.},
        title = "{X-Rays from Hybrid Stars}",
      journal = {\apj},
         year = 2005,
        month = jan,
       volume = {618},
       number = {1},
        pages = {493-501},
          doi = {10.1086/425891},
       adsurl = {https://ui.adsabs.harvard.edu/abs/2005ApJ...618..493A}
}

@ARTICLE{Ayres1981,
       author = {{Ayres}, T.~R. and {Linsky}, J.~L. and {Vaiana}, G.~S. and {Golub}, L. and {Rosner}, R.},
        title = "{The cool Half of the H-R diagram in soft X-rays.}",
      journal = {\apj},
         year = 1981,
        month = nov,
       volume = {250},
        pages = {293-299},
          doi = {10.1086/159374},
       adsurl = {https://ui.adsabs.harvard.edu/abs/1981ApJ...250..293A}
}

@ARTICLE{Haisch1991,
       author = {{Haisch}, Bernhard and {Schmitt}, J.~H.~M.~M. and {Rosso}, C.},
        title = "{The Coronal Dividing Line in the ROSAT X-Ray All-Sky Survey}",
      journal = {\apjl},
         year = 1991,
        month = dec,
       volume = {383},
        pages = {L15},
          doi = {10.1086/186230},
       adsurl = {https://ui.adsabs.harvard.edu/abs/1991ApJ...383L..15H}
}

@ARTICLE{Reimers1996,
       author = {{Reimers}, D. and {Huensch}, M. and {Schmitt}, J.~H.~M.~M. and {Toussaint}, F.},
        title = "{Hybrid stars and the reality of ``dividing lines'' among G to K bright giants and supergiants.}",
      journal = {\aap},
         year = 1996,
        month = jun,
       volume = {310},
        pages = {813-824},
       adsurl = {https://ui.adsabs.harvard.edu/abs/1996A&A...310..813R}
}

@ARTICLE{Schroeder1998,
       author = {{Schroeder}, K.-P. and {Huensch}, M. and {Schmitt}, J.~H.~M.~M.},
        title = "{X-ray activity and evolutionary status of late-type giants}",
      journal = {\aap},
         year = 1998,
        month = jul,
       volume = {335},
        pages = {591-595},
       adsurl = {https://ui.adsabs.harvard.edu/abs/1998A&A...335..591S}
}

@ARTICLE{Hunsch1998B,
       author = {{Hunsch}, M. and {Schmitt}, J.~H.~M.~M. and {Voges}, W.},
        title = "{The ROSAT all-sky survey catalogue of optically bright late-type giants and supergiants}",
      journal = {\aaps},
         year = 1998,
        month = jan,
       volume = {127},
        pages = {251-255},
          doi = {10.1051/aas:1998347},
       adsurl = {https://ui.adsabs.harvard.edu/abs/1998A&AS..127..251H}
}

@ARTICLE{Sahai2015,
       author = {{Sahai}, R. and {Sanz-Forcada}, J. and {S{\'a}nchez Contreras}, C. and {Stute}, M.},
        title = "{A Pilot Deep Survey for X-Ray Emission from fuvAGB Stars}",
      journal = {\apj},
         year = 2015,
        month = sep,
       volume = {810},
       number = {1},
          eid = {77},
        pages = {77},
          doi = {10.1088/0004-637X/810/1/77},
archivePrefix = {arXiv},
       eprint = {1507.07509},
 primaryClass = {astro-ph.SR},
       adsurl = {https://ui.adsabs.harvard.edu/abs/2015ApJ...810...77S}
}

@ARTICLE{Ortiz2021,
       author = {{Ortiz}, R. and {Guerrero}, M.~A.},
        title = "{X-Ray AGB Stars in the 4XMM-DR9 Catalog: Further Evidence for Companions}",
      journal = {\apj},
         year = 2021,
        month = may,
       volume = {912},
       number = {2},
          eid = {93},
        pages = {93},
          doi = {10.3847/1538-4357/abefd7},
       adsurl = {https://ui.adsabs.harvard.edu/abs/2021ApJ...912...93O}
}

@ARTICLE{Salvato2018,
       author = {{Salvato}, M. and {Buchner}, J. and {Budav{\'a}ri}, T. and {Dwelly}, T. and {Merloni}, A. and {Brusa}, M. and {Rau}, A. and {Fotopoulou}, S. and {Nandra}, K.},
        title = "{Finding counterparts for all-sky X-ray surveys with NWAY: a Bayesian algorithm for cross-matching multiple catalogues}",
      journal = {\mnras},
         year = 2018,
        month = feb,
       volume = {473},
       number = {4},
        pages = {4937-4955},
          doi = {10.1093/mnras/stx2651},
archivePrefix = {arXiv},
       eprint = {1705.10711},
 primaryClass = {astro-ph.GA},
       adsurl = {https://ui.adsabs.harvard.edu/abs/2018MNRAS.473.4937S}
}

@ARTICLE{Nobukawa2016,
       author = {{Nobukawa}, Masayoshi and {Uchiyama}, Hideki and {Nobukawa}, Kumiko K. and {Yamauchi}, Shigeo and {Koyama}, Katsuji},
        title = "{Origin of the Galactic Diffuse X-Ray Emission: Iron K-shell Line Diagnostics}",
      journal = {\apj},
         year = 2016,
        month = dec,
       volume = {833},
       number = {2},
          eid = {268},
        pages = {268},
          doi = {10.3847/1538-4357/833/2/268},
archivePrefix = {arXiv},
       eprint = {1701.00884},
 primaryClass = {astro-ph.HE},
       adsurl = {https://ui.adsabs.harvard.edu/abs/2016ApJ...833..268N}
}

@ARTICLE{Mukai2016,
       author = {{Mukai}, K. and {Luna}, G.~J.~M. and {Cusumano}, G. and {Segreto}, A. and {Munari}, U. and {Sokoloski}, J.~L. and {Lucy}, A.~B. and {Nelson}, T. and {Nu{\~n}ez}, N.~E.},
        title = "{SU Lyncis, a hard X-ray bright M giant: clues point to a large hidden population of symbiotic stars}",
      journal = {\mnras},
         year = 2016,
        month = sep,
       volume = {461},
       number = {1},
        pages = {L1-L5},
          doi = {10.1093/mnrasl/slw087},
archivePrefix = {arXiv},
       eprint = {1604.08483},
 primaryClass = {astro-ph.SR},
       adsurl = {https://ui.adsabs.harvard.edu/abs/2016MNRAS.461L...1M}
}

@ARTICLE{Lima2024,
       author = {{Lima}, I.~J. and {Luna}, G.~J.~M. and {Mukai}, K. and {Oliveira}, A.~S. and {Sokoloski}, J.~L. and {Walter}, F.~M. and {Palivanas}, N. and {Nu{\~n}ez}, N.~E. and {Souza}, R.~R. and {Araujo}, R.~A.~N.},
        title = "{Symbiotic stars in X-rays: IV. XMM-Newton, Swift, and TESS observations}",
      journal = {\aap},
         year = 2024,
        month = sep,
       volume = {689},
          eid = {A86},
        pages = {A86},
          doi = {10.1051/0004-6361/202449913},
archivePrefix = {arXiv},
       eprint = {2405.01508},
 primaryClass = {astro-ph.SR},
       adsurl = {https://ui.adsabs.harvard.edu/abs/2024A&A...689A..86L}
}

@ARTICLE{Ponti2026,
       author = {{Ponti}, G. and {Yeung}, M.~C.~H. and {Stel}, G. and {Locatelli}, N. and {Zheng}, X. and {Stelzer}, B. and {Merloni}, A. and {Caramazza}, M. and {Magaudda}, E. and {Sasaki}, M. and {Dennerl}, K. and {Reiprich}, T.~H. and {Schwope}, A. and {Becker}, W. and {Freyberg}, M.},
        title = "{Low-mass stars dominate the hot (0.7 keV) Galactic X-ray emission}",
      journal = {arXiv e-prints},
         year = 2026,
        month = feb,
          eid = {arXiv:2602.03941},
        pages = {arXiv:2602.03941},
          doi = {10.48550/arXiv.2602.03941},
archivePrefix = {arXiv},
       eprint = {2602.03941},
 primaryClass = {astro-ph.GA},
       adsurl = {https://ui.adsabs.harvard.edu/abs/2026arXiv260203941P}
}

@ARTICLE{Zheng2026,
       author = {{Zheng}, Xueying and {Ponti}, Gabriele and {Locatelli}, Nicola and {Stelzer}, Beate and {Magaudda}, Enza and {Dennerl}, Konrad and {Freyberg}, Michael and {Sanders}, Jeremy and {Caramazza}, Marilena and {Sasaki}, Manami and {Merloni}, Andrea and {Robrade}, Jan and {Liu}, Teng and {Zhang}, He-shou and {Mayer}, Martin G.~F. and {Zhang}, Yi and {Yeung}, Michael C.~H. and {Becker}, Werner},
        title = "{The average X-ray spectrum of the volume-complete M-, F-, G-, and K-type star sample within 10 pc of the Sun}",
      journal = {arXiv e-prints},
         year = 2026,
        month = mar,
          eid = {arXiv:2603.28751},
        pages = {arXiv:2603.28751},
          doi = {10.48550/arXiv.2603.28751},
archivePrefix = {arXiv},
       eprint = {2603.28751},
 primaryClass = {astro-ph.HE},
       adsurl = {https://ui.adsabs.harvard.edu/abs/2026arXiv260328751Z}
}

@ARTICLE{Marshall2006,
       author = {{Marshall}, D.~J. and {Robin}, A.~C. and {Reyl{\'e}}, C. and {Schultheis}, M. and {Picaud}, S.},
        title = "{Modelling the Galactic interstellar extinction distribution in three dimensions}",
      journal = {\aap},
         year = 2006,
        month = jul,
       volume = {453},
       number = {2},
        pages = {635-651},
          doi = {10.1051/0004-6361:20053842},
archivePrefix = {arXiv},
       eprint = {astro-ph/0604427},
 primaryClass = {astro-ph},
       adsurl = {https://ui.adsabs.harvard.edu/abs/2006A&A...453..635M}
}

@ARTICLE{Drew2014,
       author = {{Drew}, J.~E. and {Gonzalez-Solares}, E. and {Greimel}, R. and {Irwin}, M.~J. and {K{\"u}pc{\"u} Yoldas}, A. and {Lewis}, J. and {Barentsen}, G. and {Eisl{\"o}ffel}, J. and {Farnhill}, H.~J. and {Martin}, W.~E. and {Walsh}, J.~R. and {Walton}, N.~A. and {Mohr-Smith}, M. and {Raddi}, R. and {Sale}, S.~E. and {Wright}, N.~J. and {Groot}, P. and {Barlow}, M.~J. and {Corradi}, R.~L.~M. and {Drake}, J.~J. and {Fabregat}, J. and {Frew}, D.~J. and {G{\"a}nsicke}, B.~T. and {Knigge}, C. and {Mampaso}, A. and {Morris}, R.~A.~H. and {Naylor}, T. and {Parker}, Q.~A. and {Phillipps}, S. and {Ruhland}, C. and {Steeghs}, D. and {Unruh}, Y.~C. and {Vink}, J.~S. and {Wesson}, R. and {Zijlstra}, A.~A.},
        title = "{The VST Photometric H{\ensuremath{\alpha}} Survey of the Southern Galactic Plane and Bulge (VPHAS+)}",
      journal = {\mnras},
         year = 2014,
        month = may,
       volume = {440},
       number = {3},
        pages = {2036-2058},
          doi = {10.1093/mnras/stu394},
archivePrefix = {arXiv},
       eprint = {1402.7024},
 primaryClass = {astro-ph.GA},
       adsurl = {https://ui.adsabs.harvard.edu/abs/2014MNRAS.440.2036D}
}

@ARTICLE{Ponti2015,
       author = {{Ponti}, G. and {Morris}, M.~R. and {Terrier}, R. and {Haberl}, F. and {Sturm}, R. and {Clavel}, M. and {Soldi}, S. and {Goldwurm}, A. and {Predehl}, P. and {Nandra}, K. and {B{\'e}langer}, G. and {Warwick}, R.~S. and {Tatischeff}, V.},
        title = "{The XMM-Newton view of the central degrees of the Milky Way}",
      journal = {\mnras},
         year = 2015,
        month = oct,
       volume = {453},
       number = {1},
        pages = {172-213},
          doi = {10.1093/mnras/stv1331},
archivePrefix = {arXiv},
       eprint = {1508.04445},
 primaryClass = {astro-ph.HE},
       adsurl = {https://ui.adsabs.harvard.edu/abs/2015MNRAS.453..172P}
}

@ARTICLE{Ponti2019,
       author = {{Ponti}, G. and {Hofmann}, F. and {Churazov}, E. and {Morris}, M.~R. and {Haberl}, F. and {Nandra}, K. and {Terrier}, R. and {Clavel}, M. and {Goldwurm}, A.},
        title = "{An X-ray chimney extending hundreds of parsecs above and below the Galactic Centre}",
      journal = {\nat},
         year = 2019,
        month = mar,
       volume = {567},
       number = {7748},
        pages = {347-350},
          doi = {10.1038/s41586-019-1009-6},
archivePrefix = {arXiv},
       eprint = {1904.05969},
 primaryClass = {astro-ph.HE},
       adsurl = {https://ui.adsabs.harvard.edu/abs/2019Natur.567..347P}
}

@ARTICLE{Linsky1979,
       author = {{Linsky}, J.~L. and {Haisch}, B.~M.},
        title = "{Outer atmospheres of cool stars. I. The sharp division into solar-type and non-solar-type stars.}",
      journal = {\apjl},
         year = 1979,
        month = apr,
       volume = {229},
        pages = {L27-L32},
          doi = {10.1086/182924},
       adsurl = {https://ui.adsabs.harvard.edu/abs/1979ApJ...229L..27L}
}

@ARTICLE{Simon1982,
       author = {{Simon}, T. and {Linsky}, J.~L. and {Stencel}, R.~E.},
        title = "{On the reality of a boundary in the H-R diagram between late-type stars with and without high temperature outer atmospheres.}",
      journal = {\apj},
         year = 1982,
        month = jun,
       volume = {257},
        pages = {225-246},
          doi = {10.1086/159981},
       adsurl = {https://ui.adsabs.harvard.edu/abs/1982ApJ...257..225S}
}

@ARTICLE{Locatelli2025,
       author = {{Locatelli}, Nicola and {Ponti}, Gabriele and {Magaudda}, Enza and {Stelzer}, Beate},
        title = "{The X-ray luminosity function of stars observed with SRG/eROSITA}",
      journal = {\aap},
         year = 2025,
        month = oct,
       volume = {702},
          eid = {A237},
        pages = {A237},
          doi = {10.1051/0004-6361/202556420},
archivePrefix = {arXiv},
       eprint = {2509.22770},
 primaryClass = {astro-ph.SR},
       adsurl = {https://ui.adsabs.harvard.edu/abs/2025A&A...702A.237L}
}

@ARTICLE{Akras2023,
       author = {{Akras}, Stavros},
        title = "{Where are the missing symbiotic stars? Uncovering hidden symbiotic stars in public catalogues}",
      journal = {\mnras},
         year = 2023,
        month = mar,
       volume = {519},
       number = {4},
        pages = {6044-6054},
          doi = {10.1093/mnras/stad096},
archivePrefix = {arXiv},
       eprint = {2301.08201},
 primaryClass = {astro-ph.SR},
       adsurl = {https://ui.adsabs.harvard.edu/abs/2023MNRAS.519.6044A}
}

@ARTICLE{Xu2024,
       author = {{Xu}, Xiao-jie and {Shao}, Yong and {Li}, Xiang-Dong},
        title = "{The Missing Symbiotic Stars: A Joint Analysis with Gaia, GALEX, and XMM-Newton Data}",
      journal = {\apj},
         year = 2024,
        month = feb,
       volume = {962},
       number = {2},
          eid = {126},
        pages = {126},
          doi = {10.3847/1538-4357/ad20ec},
       adsurl = {https://ui.adsabs.harvard.edu/abs/2024ApJ...962..126X}
}

@ARTICLE{Mondal2025,
       author = {{Mondal}, Samaresh and {Ponti}, Gabriele and {Bao}, Tong and {Morris}, Mark R. and {Haberl}, Frank and {Rea}, Nanda and {Campana}, Sergio},
        title = "{A sample of ionised Fe line-emitting X-ray sources in the inner Galactic disc}",
      journal = {\aap},
         year = 2025,
        month = jun,
       volume = {698},
          eid = {A206},
        pages = {A206},
          doi = {10.1051/0004-6361/202553664},
archivePrefix = {arXiv},
       eprint = {2505.06164},
 primaryClass = {astro-ph.HE},
       adsurl = {https://ui.adsabs.harvard.edu/abs/2025A&A...698A.206M}
}

@ARTICLE{Laversveiler2025,
       author = {{Laversveiler}, Marco and {Gon{\c{c}}alves}, Denise R. and {Rocha-Pinto}, Helio J. and {Merc}, Jaroslav},
        title = "{The local group symbiotic star population and its tenuous link with type Ia supernovae}",
      journal = {\aap},
         year = 2025,
        month = jun,
       volume = {698},
          eid = {A155},
        pages = {A155},
          doi = {10.1051/0004-6361/202451548},
archivePrefix = {arXiv},
       eprint = {2504.02090},
 primaryClass = {astro-ph.GA},
       adsurl = {https://ui.adsabs.harvard.edu/abs/2025A&A...698A.155L}
}

@ARTICLE{Rodriguez2025,
       author = {{Rodriguez}, Antonio C. and {El-Badry}, Kareem and {Suleimanov}, Valery and {Pala}, Anna F. and {Kulkarni}, Shrinivas R. and {Gaensicke}, Boris and {Mori}, Kaya and {Rich}, R. Michael and {Sarkar}, Arnab and {Bao}, Tong and {Lopes de Oliveira}, Raimundo and {Ramsay}, Gavin and {Szkody}, Paula and {Graham}, Matthew and {Prince}, Thomas A. and {Caiazzo}, Ilaria and {Vanderbosch}, Zachary P. and {van Roestel}, Jan and {Das}, Kaustav K. and {Qin}, Yu-Jing and {Kasliwal}, Mansi M. and {Wold}, Avery and {Groom}, Steven L. and {Reiley}, Daniel and {Riddle}, Reed},
        title = "{Cataclysmic Variables and AM CVn Binaries in SRG/eROSITA + Gaia: Volume Limited Samples, X-Ray Luminosity Functions, and Space Densities}",
      journal = {\pasp},
         year = 2025,
        month = jan,
       volume = {137},
       number = {1},
          eid = {014201},
        pages = {014201},
          doi = {10.1088/1538-3873/ada185},
archivePrefix = {arXiv},
       eprint = {2408.16053},
 primaryClass = {astro-ph.HE},
       adsurl = {https://ui.adsabs.harvard.edu/abs/2025PASP..137a4201R}
}

@ARTICLE{Pala2020,
       author = {{Pala}, A.~F. and {G{\"a}nsicke}, B.~T. and {Breedt}, E. and {Knigge}, C. and {Hermes}, J.~J. and {Gentile Fusillo}, N.~P. and {Hollands}, M.~A. and {Naylor}, T. and {Pelisoli}, I. and {Schreiber}, M.~R. and {Toonen}, S. and {Aungwerojwit}, A. and {Cukanovaite}, E. and {Dennihy}, E. and {Manser}, C.~J. and {Pretorius}, M.~L. and {Scaringi}, S. and {Toloza}, O.},
        title = "{A Volume-limited Sample of Cataclysmic Variables from Gaia DR2: Space Density and Population Properties}",
      journal = {\mnras},
         year = 2020,
        month = may,
       volume = {494},
       number = {3},
        pages = {3799-3827},
          doi = {10.1093/mnras/staa764},
archivePrefix = {arXiv},
       eprint = {1907.13152},
 primaryClass = {astro-ph.SR},
       adsurl = {https://ui.adsabs.harvard.edu/abs/2020MNRAS.494.3799P}
}

@ARTICLE{Belloni2018,
       author = {{Belloni}, Diogo and {Schreiber}, Matthias R. and {Zorotovic}, M{\'o}nica and {I{\l}kiewicz}, Krystian and {Hurley}, Jarrod R. and {Giersz}, Mirek and {Lagos}, Felipe},
        title = "{No cataclysmic variables missing: higher merger rate brings into agreement observed and predicted space densities}",
      journal = {\mnras},
         year = 2018,
        month = aug,
       volume = {478},
       number = {4},
        pages = {5626-5637},
          doi = {10.1093/mnras/sty1421},
archivePrefix = {arXiv},
       eprint = {1806.05687},
 primaryClass = {astro-ph.SR},
       adsurl = {https://ui.adsabs.harvard.edu/abs/2018MNRAS.478.5626B}
}

@ARTICLE{Revnivtsev2008,
       author = {{Revnivtsev}, M. and {Sazonov}, S. and {Krivonos}, R. and {Ritter}, H. and {Sunyaev}, R.},
        title = "{Properties of the Galactic population of cataclysmic variables in hard X-rays}",
      journal = {\aap},
         year = 2008,
        month = oct,
       volume = {489},
       number = {3},
        pages = {1121-1127},
          doi = {10.1051/0004-6361:200810213},
archivePrefix = {arXiv},
       eprint = {0805.2699},
 primaryClass = {astro-ph},
       adsurl = {https://ui.adsabs.harvard.edu/abs/2008A&A...489.1121R}
}

@ARTICLE{Pretorius2012,
       author = {{Pretorius}, Magaretha L. and {Knigge}, Christian},
        title = "{The space density and X-ray luminosity function of non-magnetic cataclysmic variables}",
      journal = {\mnras},
         year = 2012,
        month = jan,
       volume = {419},
       number = {2},
        pages = {1442-1454},
          doi = {10.1111/j.1365-2966.2011.19801.x},
archivePrefix = {arXiv},
       eprint = {1109.3162},
 primaryClass = {astro-ph.SR},
       adsurl = {https://ui.adsabs.harvard.edu/abs/2012MNRAS.419.1442P}
}

@ARTICLE{Pretorius2013,
       author = {{Pretorius}, Magaretha L. and {Knigge}, Christian and {Schwope}, Axel D.},
        title = "{The space density of magnetic cataclysmic variables}",
      journal = {\mnras},
         year = 2013,
        month = jun,
       volume = {432},
       number = {1},
        pages = {570-583},
          doi = {10.1093/mnras/stt499},
archivePrefix = {arXiv},
       eprint = {1303.4270},
 primaryClass = {astro-ph.SR},
       adsurl = {https://ui.adsabs.harvard.edu/abs/2013MNRAS.432..570P}
}

@ARTICLE{Yu2022,
       author = {{Yu}, Zhuo-li and {Xu}, Xiao-jie and {Shao}, Yong and {Wang}, Q. Daniel and {Li}, Xiang-Dong},
        title = "{Y Gem: A White Dwarf Symbiotic Star?}",
      journal = {\apj},
         year = 2022,
        month = jun,
       volume = {932},
       number = {2},
          eid = {132},
        pages = {132},
          doi = {10.3847/1538-4357/ac6ba0},
       adsurl = {https://ui.adsabs.harvard.edu/abs/2022ApJ...932..132Y}
}

@ARTICLE{Meng2016,
       author = {{Meng}, Xiangcun and {Han}, Zhanwen},
        title = "{The X-ray/radio and UV luminosity expected from symbiotic systems as the progenitor of SNe Ia}",
      journal = {\aap},
         year = 2016,
        month = apr,
       volume = {588},
          eid = {A88},
        pages = {A88},
          doi = {10.1051/0004-6361/201526077},
archivePrefix = {arXiv},
       eprint = {1601.07274},
 primaryClass = {astro-ph.SR},
       adsurl = {https://ui.adsabs.harvard.edu/abs/2016A&A...588A..88M}
}

@ARTICLE{Guerrero2024,
       author = {{Guerrero}, M.~A. and {Montez}, R. and {Ortiz}, R. and {Toal{\'a}}, J.~A. and {Kastner}, J.~H.},
        title = "{Asymptotic giant branch stars in the eROSITA-DE eRASS1 catalog}",
      journal = {\aap},
         year = 2024,
        month = sep,
       volume = {689},
          eid = {A62},
        pages = {A62},
          doi = {10.1051/0004-6361/202450155},
archivePrefix = {arXiv},
       eprint = {2407.10552},
 primaryClass = {astro-ph.SR},
       adsurl = {https://ui.adsabs.harvard.edu/abs/2024A&A...689A..62G}
}

@ARTICLE{Binney1997,
       author = {{Binney}, James and {Gerhard}, Ortwin and {Spergel}, David},
        title = "{The photometric structure of the inner Galaxy}",
      journal = {\mnras},
         year = 1997,
        month = jun,
       volume = {288},
       number = {2},
        pages = {365-374},
          doi = {10.1093/mnras/288.2.365},
archivePrefix = {arXiv},
       eprint = {astro-ph/9609066},
 primaryClass = {astro-ph},
       adsurl = {https://ui.adsabs.harvard.edu/abs/1997MNRAS.288..365B}
}

@ARTICLE{Munari2019,
       author = {{Munari}, Ulisse},
        title = "{The Symbiotic Stars}",
      journal = {arXiv e-prints},
         year = 2019,
        month = sep,
          eid = {arXiv:1909.01389},
        pages = {arXiv:1909.01389},
          doi = {10.48550/arXiv.1909.01389},
archivePrefix = {arXiv},
       eprint = {1909.01389},
 primaryClass = {astro-ph.SR},
       adsurl = {https://ui.adsabs.harvard.edu/abs/2019arXiv190901389M}
}

@ARTICLE{Bondi1944,
       author = {{Bondi}, H. and {Hoyle}, F.},
        title = "{On the mechanism of accretion by stars}",
      journal = {\mnras},
         year = 1944,
        month = jan,
       volume = {104},
        pages = {273},
          doi = {10.1093/mnras/104.5.273},
       adsurl = {https://ui.adsabs.harvard.edu/abs/1944MNRAS.104..273B}
}

@ARTICLE{Podsiadlowski2007,
       author = {{Podsiadlowski}, Ph. and {Mohamed}, S.},
        title = "{The Origin and Evolution of Symbiotic Binaries}",
      journal = {Baltic Astronomy},
         year = 2007,
        month = jan,
       volume = {16},
        pages = {26-33},
       adsurl = {https://ui.adsabs.harvard.edu/abs/2007BaltA..16...26P}
}

@ARTICLE{Eze2014,
       author = {{Eze}, R.~N.~C.},
        title = "{Fe K{\ensuremath{\alpha}} line in hard X-ray emitting symbiotic stars}",
      journal = {\mnras},
         year = 2014,
        month = jan,
       volume = {437},
       number = {1},
        pages = {857-861},
          doi = {10.1093/mnras/stt1947},
archivePrefix = {arXiv},
       eprint = {1310.2461},
 primaryClass = {astro-ph.HE},
       adsurl = {https://ui.adsabs.harvard.edu/abs/2014MNRAS.437..857E}
}

@ARTICLE{Toala2024,
       author = {{Toal{\'a}}, Jes{\'u}s A.},
        title = "{Reflection physics in X-ray-emitting symbiotic stars}",
      journal = {\mnras},
         year = 2024,
        month = feb,
       volume = {528},
       number = {1},
        pages = {987-996},
          doi = {10.1093/mnras/stae039},
archivePrefix = {arXiv},
       eprint = {2401.02318},
 primaryClass = {astro-ph.HE},
       adsurl = {https://ui.adsabs.harvard.edu/abs/2024MNRAS.528..987T}
}

@ARTICLE{Barnbaum1995,
       author = {{Barnbaum}, Cecilia and {Morris}, Mark and {Kahane}, Claudine},
        title = "{Evidence for Rapid Rotation of the Carbon Star V Hydrae}",
      journal = {\apj},
         year = 1995,
        month = sep,
       volume = {450},
        pages = {862},
          doi = {10.1086/176190},
       adsurl = {https://ui.adsabs.harvard.edu/abs/1995ApJ...450..862B}
}

@BOOK{Kenyon1986,
       author = {{Kenyon}, S.~J.},
        title = "{The symbiotic stars}",
         year = 1986,
       adsurl = {https://ui.adsabs.harvard.edu/abs/1986syst.book.....K}
}

@INPROCEEDINGS{Magrini2003,
       author = {{Magrini}, L. and {Corradi}, R.~L.~M. and {Munari}, U.},
        title = "{A Search for Symbiotic Stars in the Local Group}",
    booktitle = {Symbiotic Stars Probing Stellar Evolution},
         year = 2003,
       editor = {{Corradi}, R.~L.~M. and {Mikolajewska}, J. and {Mahoney}, T.~J.},
       series = {Astronomical Society of the Pacific Conference Series},
       volume = {303},
        month = jan,
        pages = {539},
          doi = {10.48550/arXiv.astro-ph/0208085},
archivePrefix = {arXiv},
       eprint = {astro-ph/0208085},
 primaryClass = {astro-ph},
       adsurl = {https://ui.adsabs.harvard.edu/abs/2003ASPC..303..539M}
}

@ARTICLE{NewAthena,
       author = {{Cruise}, Mike and {Guainazzi}, Matteo and {Aird}, James and {Carrera}, Francisco J. and {Costantini}, Elisa and {Corrales}, Lia and {Dauser}, Thomas and {Eckert}, Dominique and {Gastaldello}, Fabio and {Matsumoto}, Hironori and {Osten}, Rachel and {Petrucci}, Pierre-Olivier and {Porquet}, Delphine and {Pratt}, Gabriel W. and {Rea}, Nanda and {Reiprich}, Thomas H. and {Simionescu}, Aurora and {Spiga}, Daniele and {Troja}, Eleonora},
        title = "{The NewAthena mission concept in the context of the next decade of X-ray astronomy}",
      journal = {Nature Astronomy},
         year = 2025,
        month = jan,
       volume = {9},
        pages = {36-44},
          doi = {10.1038/s41550-024-02416-3},
archivePrefix = {arXiv},
       eprint = {2501.03100},
 primaryClass = {astro-ph.IM},
       adsurl = {https://ui.adsabs.harvard.edu/abs/2025NatAs...9...36C}
}

%%%%%%%%%%%%%%%%%%%%%%%%%%%%%%%%%%%%%%%%%%%%%%%%%%%%%%%%%%%%%%%
% Appendices must be placed after   \end{thebibliography}
% They will be placed automatically on a new page.
%%%%%%%%%%%%%%%%%%%%%%%%%%%%%%%%%%%%%%%%%%%%%%%%%%%%%%%%%%%%%%%
\begin{appendix}
\onecolumn
\section{Symbiotic star candidates}
\setlength{\tabcolsep}{3pt}
\renewcommand{\arraystretch}{1.5}
\begin{longtable}{llllllllllllcc}
\caption{\raggedright \label{tab:specsrc} Source properties for 107 SySt candidates (The full table would be available at the CDS).}\\

\toprule
ID & RA & DEC & Sep & G (mag) & BP-RP & distance  & $\rm d_{max, X}$ & $\rm d_{max, G}$  & $\log_{10}L_{\rm X}$ & $\rm p_{any}$ & $\rm p_{i}$  & $\rm (r-H\alpha)_v$ & $\rm (u-i)_v$ \\
 &(deg) & (deg) & (arcsec) &  &  & (kpc)  & (kpc)  & (kpc)   & $\rm erg~s^{-1}$ &  &   &  &  \\
\midrule
\endfirsthead

\midrule
\endhead
\bottomrule
\endfoot
1   & 265.1487 & -28.0949 & 0.20 & 14.45 & 3.29 & 7.70 & 15.40 & 20.00 & 32.15 & 0.98 & 1.00 & 0.60 & 6.61 \\
2   & 265.9368 & -28.0362 & 0.58 & 16.93 & 4.93 & 7.78 & 20.00 & 3.16 & 33.17 & 0.89 & 1.00 & 0.66 & -- \\
3   & 265.9693 & -27.5573 & 0.97 & 13.96 & 2.49 & 2.40 & 6.12 & 7.93 & 31.35 & 0.93 & 1.00 & 0.54 & 5.56 \\
4   & 266.5127 & -27.9757 & 0.39 & 12.61 & 2.30 & 1.15 & 2.54 & 2.40 & 30.59 & 0.95 & 1.00 & 0.48 & 5.75 \\
5   & 266.8106 & -28.6865 & 0.13 & 16.83 & 3.13 & 1.72 & 2.95 & 2.26 & 30.72 & 0.97 & 1.00 & 0.65 & -- \\
6   & 266.8204 & -28.7873 & 0.18 & 11.23 & 2.13 & 2.34 & 5.55 & 3.13 & 31.27 & 0.99 & 1.00 & 0.53 & 3.33 \\
7   & 266.8668 & -28.0730 & 0.72 & 17.74 & 3.46 & 1.72 & 1.94 & 2.25 & 30.35 & 0.86 & 0.58 & 0.59 & -- \\
8   & 266.8925 & -28.4018 & 1.19 & 11.83 & 3.25 & 0.84 & 1.28 & 2.45 & 29.99 & 0.86 & 1.00 & 0.74 & 7.08 \\
9   & 267.0736 & -28.9131 & 0.46 & 16.42 & 6.52 & 3.11 & 20.00 & 3.64 & 32.50 & 0.91 & 1.00 & 0.61 & -- \\
10   & 267.2121 & -28.2461 & 0.45 & 7.83 & 1.57 & 0.91 & 1.92 & 3.83 & 30.34 & 0.96 & 1.00 & $^\dag$1.21 & 2.53 \\
11   & 267.2678 & -28.4906 & 1.12 & 14.06 & 2.96 & 2.45 & 3.01 & 4.22 & 30.73 & 0.92 & 1.00 & 0.62 & 6.95 \\
12   & 267.3310 & -28.4526 & 0.62 & 13.75 & 3.74 & 5.17 & 9.64 & 4.97 & 31.75 & 0.94 & 1.00 & 0.65 & 7.96 \\
13   & 267.3492 & -28.7536 & 1.23 & 14.69 & 3.98 & 4.56 & 7.25 & 5.31 & 31.50 & 0.90 & 1.00 & 0.40 & -- \\
14   & 267.5009 & -29.0640 & 1.47 & 12.16 & 2.88 & 3.17 & 5.55 & 9.04 & 31.27 & 0.88 & 0.99 & 0.60 & 5.51 \\
15   & 267.8372 & -28.7320 & 0.26 & 12.83 & 2.04 & 1.26 & 6.23 & 5.69 & 31.37 & 0.99 & 1.00 & 0.54 & 4.41 \\
16   & 268.2296 & -29.5731 & 1.85 & 14.13 & 3.39 & 2.74 & 5.56 & 20.00 & 31.27 & 0.88 & 0.62 & $^\dag$0.82 & -- \\
17   & 268.4744 & -29.8781 & 1.55 & 14.24 & 1.88 & 2.13 & 2.73 & 20.00 & 30.65 & 0.89 & 0.77 & 0.48 & 4.23 \\
18   & 268.5203 & -29.8376 & 1.63 & 13.18 & 3.61 & 3.11 & 4.56 & 20.00 & 31.09 & 0.86 & 0.92 & 0.73 & 8.40 \\
19   & 268.6645 & -29.4797 & 1.10 & 13.75 & 2.84 & 3.61 & 6.44 & 20.00 & 31.39 & 0.90 & 0.81 & 0.59 & -- \\
20   & 267.0820 & -28.1241 & 0.59 & 11.87 & 5.97 & 2.14 & 10.06 & 4.27 & 31.78 & 0.86 & 1.00 & 0.77 & 8.89 \\
21   & 267.1494 & -27.9384 & 0.08 & 15.17 & 5.32 & 1.70 & 8.94 & 3.39 & 31.68 & 1.00 & 1.00 & 0.70 & -- \\
22   & 268.1189 & -27.3713 & 0.79 & 14.50 & 3.24 & 2.10 & 10.73 & 3.66 & 31.84 & 0.96 & 1.00 & 0.59 & 7.43 \\
23   & 268.7641 & -26.7172 & 0.50 & 16.44 & 4.12 & 0.90 & 4.39 & 2.00 & 31.06 & 0.93 & 1.00 & 0.60 & -- \\
24   & 268.0393 & -26.0776 & 2.97 & 11.12 & 3.27 & 1.39 & 2.72 & 3.68 & 30.65 & 0.88 & 0.81 & 0.69 & 7.26 \\
25   & 268.2797 & -26.3802 & 2.06 & 12.47 & 2.82 & 1.00 & 1.43 & 3.30 & 30.09 & 0.89 & 0.96 & 0.54 & 5.25 \\
26   & 268.4682 & -25.5932 & 1.28 & 15.77 & 3.90 & $^\ddag$6.18 & 18.35 & 4.75 & 32.30 & 0.88 & 1.00 & 0.64 & -- \\
27   & 268.4963 & -25.5574 & 1.20 & 16.39 & 3.30 & 1.46 & 3.94 & 2.84 & 30.97 & 0.82 & 0.78 & 0.61 & -- \\
28   & 268.9489 & -26.3173 & 0.64 & 11.94 & 2.40 & 2.40 & 10.48 & 4.59 & 31.82 & 0.95 & 0.94 & 0.48 & 5.69 \\
29   & 269.2859 & -25.8552 & 1.22 & 14.90 & 6.71 & 5.82 & 20.00 & 6.54 & 32.39 & 0.94 & 0.85 & $^\dag$0.84 & -- \\
30   & 267.9519 & -25.0228 & 0.45 & 16.63 & 3.61 & 1.38 & 3.39 & 3.05 & 30.84 & 0.87 & 0.91 & 0.61 & -- \\
31   & 268.3541 & -24.8560 & 0.55 & 16.22 & 3.53 & 4.36 & 8.71 & 4.08 & 31.66 & 0.93 & 1.00 & 0.60 & -- \\
32   & 268.6278 & -25.2189 & 2.34 & 11.82 & 5.53 & 2.46 & 3.95 & 5.36 & 30.97 & 0.96 & 0.94 & 0.76 & 8.63 \\
33   & 268.8282 & -23.8852 & 0.68 & 15.67 & 3.95 & 2.47 & 6.75 & 3.34 & 31.44 & 0.94 & 1.00 & 0.68 & -- \\
34   & 270.0488 & -24.0747 & 0.37 & 13.00 & 5.08 & 2.83 & 3.67 & 4.23 & 30.91 & 0.99 & 1.00 & 0.75 & 8.52 \\
35   & 270.2504 & -24.4459 & 1.16 & 13.52 & 2.73 & 4.42 & 12.75 & 4.61 & 31.99 & 0.97 & 1.00 & 0.55 & 6.68 \\
36   & 270.1431 & -22.7944 & 0.40 & 12.60 & 3.21 & 3.29 & 20.00 & 4.86 & 32.65 & 1.00 & 1.00 & $^\dag$0.80 & -- \\
37   & 270.3551 & -23.6152 & 0.59 & 10.98 & 2.35 & 3.65 & 4.38 & 4.98 & 31.06 & 0.99 & 1.00 & 0.06 & -- \\
38   & 270.3585 & -24.0195 & 0.85 & 17.04 & 6.07 & 1.44 & 13.89 & 2.94 & 32.06 & 0.91 & 1.00 & 0.79 & -- \\
39   & 270.3664 & -23.7452 & 2.50 & 12.20 & 2.98 & 3.67 & 6.15 & 4.74 & 31.35 & 0.98 & 1.00 & 0.46 & 5.82 \\
40   & 270.4443 & -23.5776 & 0.88 & 12.47 & 3.29 & 2.84 & 3.40 & 4.46 & 30.84 & 0.99 & 1.00 & 0.59 & 6.72 \\
41   & 270.5054 & -23.7982 & 0.18 & 10.92 & 2.76 & 3.50 & 9.70 & 5.00 & 31.75 & 1.00 & 1.00 & 0.45 & 5.82 \\
42   & 270.5171 & -23.6282 & 0.62 & 12.90 & 3.46 & 0.64 & 5.29 & 3.34 & 31.22 & 0.99 & 1.00 & $^\dag$0.92 & 5.27 \\
43   & 270.5440 & -24.0139 & 0.80 & 9.89 & 3.55 & 1.43 & 3.14 & 5.07 & 30.77 & 0.98 & 1.00 & 0.10 & 7.49 \\
44   & 270.5553 & -23.5963 & 1.15 & 12.06 & 2.65 & 4.21 & 7.11 & 4.82 & 31.48 & 0.99 & 1.00 & 0.56 & 4.88 \\
45   & 270.5980 & -23.5772 & 0.91 & 10.59 & 2.96 & 5.31 & 20.00 & 7.03 & 33.14 & 0.99 & 1.00 & $^\dag$0.97 & 6.30 \\
46   & 269.2109 & -21.8169 & 0.15 & 17.31 & 3.43 & 1.02 & 2.33 & 1.95 & 30.51 & 0.96 & 1.00 & $^\dag$0.81 & -- \\
47   & 269.3166 & -21.6913 & 0.45 & 11.87 & 4.96 & 1.48 & 1.74 & 7.54 & 30.26 & 0.99 & 1.00 & $^\dag$1.75 & 6.98 \\
48   & 260.6915 & -37.5014 & 4.48 & 12.18 & 3.92 & 1.96 & 2.34 & 4.80 & 30.51 & 0.94 & 0.55 & 0.76 & -- \\
49   & 260.1264 & -35.8799 & 1.56 & 15.27 & 3.78 & 3.51 & 6.37 & 3.35 & 31.39 & 0.93 & 0.98 & 0.59 & 5.75 \\
50   & 260.7122 & -36.0402 & 1.37 & 15.98 & 3.81 & 1.23 & 4.64 & 2.46 & 31.11 & 0.94 & 0.92 & 0.62 & -- \\
51   & 260.8534 & -35.8678 & 0.76 & 16.46 & 2.85 & 1.31 & 4.81 & 2.34 & 31.14 & 0.98 & 1.00 & 0.61 & -- \\
52   & 261.4481 & -36.2746 & 0.71 & 15.74 & 2.88 & 1.47 & 2.50 & 3.14 & 30.57 & 0.98 & 1.00 & 0.61 & -- \\
53   & 261.5267 & -36.3698 & 0.92 & 17.01 & 3.21 & 1.11 & 2.98 & 1.73 & 30.73 & 0.94 & 1.00 & 0.66 & -- \\
54   & 260.9989 & -35.2252 & 1.42 & 16.87 & 3.05 & 1.47 & 2.62 & 1.78 & 30.61 & 0.91 & 1.00 & 0.59 & -- \\
55   & 261.0412 & -35.2398 & 1.35 & 15.93 & 2.99 & 1.09 & 3.96 & 1.71 & 30.97 & 0.94 & 1.00 & 0.58 & -- \\
56   & 261.0915 & -35.3552 & 0.91 & 12.93 & 2.61 & 0.76 & 1.47 & 1.98 & 30.11 & 1.00 & 0.64 & 0.54 & 4.55 \\
57   & 261.6954 & -35.0491 & 1.94 & 14.28 & 4.77 & 1.03 & 1.01 & 2.10 & 29.78 & 0.94 & 1.00 & 0.08 & -- \\
58   & 261.7820 & -35.7637 & 1.43 & 15.20 & 4.89 & 1.28 & 2.51 & 2.64 & 30.58 & 0.92 & 0.99 & 0.63 & -- \\
59   & 262.6367 & -35.4209 & 0.55 & 16.78 & 3.49 & 1.25 & 3.80 & 2.37 & 30.94 & 0.92 & 0.93 & 0.60 & -- \\
60   & 261.2718 & -34.3501 & 2.38 & 15.06 & 3.04 & 1.40 & 2.07 & 2.34 & 30.41 & 0.89 & 1.00 & 0.50 & 7.02 \\
61   & 261.6849 & -34.5950 & 0.57 & 11.15 & 2.80 & 1.76 & 10.55 & 3.58 & 31.82 & 1.00 & 1.00 & -0.07 & 4.95 \\
62   & 262.7230 & -34.3664 & 1.76 & 13.12 & 3.24 & 3.75 & 6.05 & 3.62 & 31.34 & 0.95 & 1.00 & 0.59 & 6.08 \\
63   & 263.2723 & -34.8906 & 2.06 & 13.95 & 2.59 & 0.91 & 0.74 & 3.58 & 29.51 & 0.88 & 0.93 & 0.49 & 5.52 \\
64   & 262.2541 & -31.9676 & 0.79 & 16.09 & 6.43 & 4.15 & 20.00 & 5.12 & 32.54 & 0.85 & 0.95 & $^\dag$0.80 & -- \\
65   & 263.0680 & -32.3151 & 0.30 & 15.37 & 3.43 & 4.60 & 9.01 & 5.01 & 31.69 & 0.96 & 0.95 & 0.66 & -- \\
66   & 263.1180 & -32.3446 & 0.55 & 11.86 & 2.48 & 2.96 & 4.52 & 5.89 & 31.09 & 0.98 & 1.00 & 0.53 & 4.31 \\
67   & 263.4421 & -33.1193 & 1.01 & 8.94 & 4.68 & 1.88 & 17.56 & 11.65 & 32.27 & 0.99 & 1.00 & 0.66 & 8.15 \\
68   & 264.0748 & -33.1222 & 1.01 & 13.29 & 4.88 & 7.30 & 20.00 & 10.85 & 32.82 & 0.97 & 1.00 & $^\dag$0.87 & 8.45 \\
69   & 264.1136 & -33.4933 & 1.71 & 9.31 & 2.06 & 0.97 & 2.53 & 19.22 & 30.58 & 0.98 & 1.00 & 0.76 & 3.82 \\
70   & 264.2083 & -32.7753 & 0.75 & 16.25 & 3.47 & 1.53 & 3.31 & 3.23 & 30.82 & 0.93 & 0.90 & 0.66 & -- \\
71   & 264.4502 & -33.3696 & 0.66 & 15.98 & 3.94 & 1.14 & 3.37 & 3.32 & 30.83 & 0.95 & 1.00 & 0.65 & -- \\
72   & 263.2851 & -31.9632 & 0.12 & 16.52 & 3.38 & 1.83 & 3.44 & 3.02 & 30.85 & 0.89 & 1.00 & 0.53 & -- \\
73   & 263.4792 & -31.6601 & 0.64 & 15.36 & 4.19 & 1.49 & 3.72 & 2.64 & 30.92 & 0.94 & 1.00 & 0.65 & -- \\
74   & 263.9960 & -31.1917 & 0.84 & 15.97 & 4.15 & 1.27 & 2.24 & 2.53 & 30.48 & 0.94 & 1.00 & 0.65 & -- \\
75   & 264.3671 & -31.5157 & 0.29 & 12.50 & 2.80 & 1.71 & 14.49 & 4.12 & 32.10 & 1.00 & 1.00 & 0.57 & 6.48 \\
76   & 264.7175 & -32.4653 & 0.51 & 14.83 & 2.84 & 1.20 & 3.10 & 2.93 & 30.76 & 0.99 & 1.00 & 0.57 & 6.00 \\
77   & 264.7447 & -32.7095 & 1.24 & 15.36 & 2.97 & 1.47 & 4.17 & 3.53 & 31.02 & 0.90 & 1.00 & 0.49 & -- \\
78   & 264.9032 & -31.8615 & 1.26 & 14.90 & 2.79 & 1.24 & 4.50 & 2.49 & 31.08 & 0.91 & 1.00 & 0.54 & 6.22 \\
79   & 264.9934 & -31.9564 & 0.17 & 16.71 & 5.15 & 4.38 & 12.82 & 3.96 & 31.99 & 0.96 & 1.00 & 0.63 & -- \\
80   & 265.2960 & -32.3017 & 1.13 & 14.51 & 4.39 & 3.22 & 5.13 & 7.16 & 31.20 & 0.94 & 0.77 & $^\dag$0.80 & -- \\
81   & 263.4637 & -30.3264 & 1.45 & 14.24 & 5.81 & 1.59 & 7.04 & 6.59 & 31.47 & 0.96 & 1.00 & $^\dag$1.11 & -- \\
82   & 264.2747 & -31.1570 & 0.71 & 13.47 & 2.39 & 1.72 & 2.57 & 3.27 & 30.60 & 0.97 & 1.00 & 0.52 & 5.61 \\
83   & 264.9432 & -31.6003 & 0.41 & 16.64 & 4.75 & 2.99 & 7.69 & 3.14 & 31.55 & 0.98 & 1.00 & 0.66 & -- \\
84   & 264.9838 & -31.1599 & 1.01 & 15.35 & 4.58 & 2.38 & 7.13 & 3.31 & 31.48 & 0.96 & 0.99 & $^\dag$0.80 & -- \\
85   & 265.0984 & -30.9867 & 1.11 & 14.12 & 5.46 & 7.24 & 6.64 & 4.52 & 31.42 & 0.95 & 1.00 & 0.79 & -- \\
86   & 265.5156 & -30.8494 & 0.30 & 14.81 & 2.58 & 1.05 & 4.68 & 2.25 & 31.12 & 1.00 & 1.00 & 0.57 & 5.68 \\
87   & 264.7573 & -29.8287 & 1.34 & 12.68 & 2.76 & 3.25 & 3.86 & 6.47 & 30.95 & 0.97 & 1.00 & 0.57 & 5.42 \\
88   & 264.7737 & -30.2358 & 1.23 & 12.05 & 3.37 & 2.66 & 3.70 & 5.28 & 30.91 & 0.99 & 1.00 & 0.75 & 6.87 \\
89   & 264.8066 & -29.5920 & 0.37 & 16.82 & 4.06 & 1.08 & 2.06 & 2.03 & 30.40 & 0.87 & 1.00 & 0.70 & -- \\
90   & 264.9409 & -29.5428 & 0.62 & 17.92 & 5.03 & 0.53 & 1.70 & 0.91 & 30.24 & 0.89 & 0.89 & $^\dag$0.80 & -- \\
91   & 265.2755 & -30.4326 & 0.50 & 17.30 & 2.86 & 2.59 & 6.70 & 2.31 & 31.43 & 0.90 & 0.63 & 0.56 & -- \\
92   & 265.6672 & -30.0121 & 0.68 & 15.64 & 5.07 & 1.66 & 0.95 & 2.44 & 29.73 & 0.94 & 1.00 & 0.65 & -- \\
93   & 266.2150 & -30.7779 & 0.38 & 16.95 & 4.12 & 1.38 & 3.74 & 2.32 & 30.92 & 0.97 & 1.00 & 0.60 & -- \\
94   & 266.2902 & -31.0442 & 0.81 & 14.46 & 3.73 & 2.70 & 6.10 & 4.54 & 31.35 & 0.97 & 1.00 & 0.61 & -- \\
95   & 266.6382 & -31.3706 & 0.37 & 16.75 & 3.68 & 1.52 & 6.07 & 2.84 & 31.34 & 0.97 & 1.00 & 0.68 & -- \\
96   & 267.0476 & -30.4605 & 0.94 & 14.13 & 2.76 & 8.51 & 20.00 & 20.00 & 32.82 & 0.97 & 1.00 & 0.58 & 5.78 \\
97   & 267.7897 & -31.1390 & 1.04 & 16.38 & 3.69 & 1.88 & 8.36 & 3.29 & 31.62 & 0.90 & 0.95 & 0.66 & -- \\
98   & 267.8568 & -31.1147 & 1.71 & 14.33 & 5.45 & 3.78 & 10.23 & 18.84 & 31.80 & 0.92 & 0.99 & $^\dag$0.91 & -- \\
99   & 264.8084 & -29.1680 & 0.86 & 13.71 & 6.15 & 4.51 & 20.00 & 10.22 & 32.83 & 0.97 & 1.00 & $^\dag$0.83 & -- \\
100   & 265.3946 & -29.1454 & 0.56 & 13.84 & 4.21 & 3.52 & 5.18 & 2.30 & 31.21 & 0.90 & 0.97 & 0.62 & 8.30 \\
101   & 265.5742 & -28.9466 & 0.29 & 11.36 & 2.97 & 5.24 & 20.00 & 4.07 & 32.63 & 1.00 & 1.00 & 0.67 & 6.08 \\
102   & 265.8309 & -29.2333 & 0.12 & 14.35 & 4.19 & 3.44 & 20.00 & 2.50 & 32.38 & 0.99 & 1.00 & 0.68 & -- \\
103   & 265.8905 & -28.9239 & 0.54 & 16.33 & 4.49 & 1.03 & 2.60 & 1.64 & 30.61 & 0.89 & 1.00 & 0.71 & -- \\
104   & 266.0975 & -29.2950 & 0.17 & 17.36 & 3.16 & 1.73 & 5.63 & 1.59 & 31.28 & 0.98 & 1.00 & 0.62 & -- \\
105   & 266.5043 & -30.1370 & 0.38 & 13.85 & 2.47 & 3.17 & 7.52 & 3.26 & 31.53 & 0.97 & 1.00 & 0.50 & 5.22 \\
106   & 267.6445 & -29.9776 & 2.09 & 12.17 & 2.06 & 4.13 & 3.46 & 20.00 & 30.86 & 0.89 & 0.95 & 0.42 & 5.26 \\
107   & 267.9628 & -30.0009 & 0.54 & 14.89 & 3.57 & 3.67 & 6.74 & 19.27 & 31.43 & 0.90 & 0.90 & 0.62 & -- \\

\end{longtable}
\tablefoot{Column(1)-(3): Source indices and coordinates from X-ray; 
Column (4): Angular separation between the X-ray source and the Gaia counterpart; 
Columns (5)-(6): Gaia G-band magnitude and BP-RP color;  
Column (7): Geometric distance from Gaia DR3;
Columns (8)--(9): Maximum detectable distance based on X-ray and Gaia G-band sensitivity, respectively; 
Column (10): Logarithm of X-ray luminosity in the 0.2--12 keV band; 
Column (11): For each entry in the X-ray catalogue, the probability that any of the associations is the correct one;
Column (12): Relative probability of the match, if one exists. The p\_i add up to unity for each X-ray source;
Columns (13)--(14): VPHAS+ survey derived colors after extinction correction. Potential $\rm H{\alpha}$ excess with (r-$\rm H{\alpha}$) >0.8  are labeled by $\dag$. \\}
\section{Comprehensive X-ray spectral analysis of SySt candidates}
\label{app:spectra}
The identification of SySts typically relies on the detection of Asymptotic Giant Branch (AGB) stars, which are considered the primary optical companions in such systems. According to the selection criteria described by \citet{Akras2023} and \citet{Xu2024}, the majority of identified AGB stars occupies an absolute magnitude range of $-5 < M_G < 5$. Here we divided our sample of 107 SySt candidates into two subgroups: more-luminous and less-luminous by applying a magnitude cut at $M_G = 3.0$, where potential YSO contamination begins to emerge, resulting in 76 and 31 sources, respectively. 
This stratification was performed to verify that the prominent EW of the 6.7~keV line is an intrinsic feature of the population rather than an effect biased toward the optically luminous subsample. As indicated by the color scale in Fig.~\ref{fig:HR_LX}, higher $M_G$ values (lower optical luminosity) generally correspond to lower X-ray luminosities. Furthermore, the classification in Fig.~\ref{fig:CMD_gaiaclass} suggests that the less-luminous group may be more susceptible to contamination from YSOs.

By analyzing these subsamples separately, we ensure that the properties derived from the stacked spectra are not biased or dominated by the most luminous sources. As shown in Fig.~\ref{fig:LPV2spec}, the black and red lines represent the stacked spectra of the more-luminous and less-luminous samples, respectively. The measured EWs for the 6.7~keV line in these two groups are $930 \pm 170$~eV and $764 \pm 330$~eV. These consistent measurements confirm that the high-ionization iron emission is a robust characteristic across different luminosity regimes within our candidate sample. Moreover, the spectral similarity between the two groups reinforces the reliability of our classification, suggesting that the less-luminous sources are not misidentified YSOs but share the same X-ray spectral nature as the primary LPV sample.
\begin{figure*}[h]
  \centering
  
  \includegraphics[width=1.0\hsize]{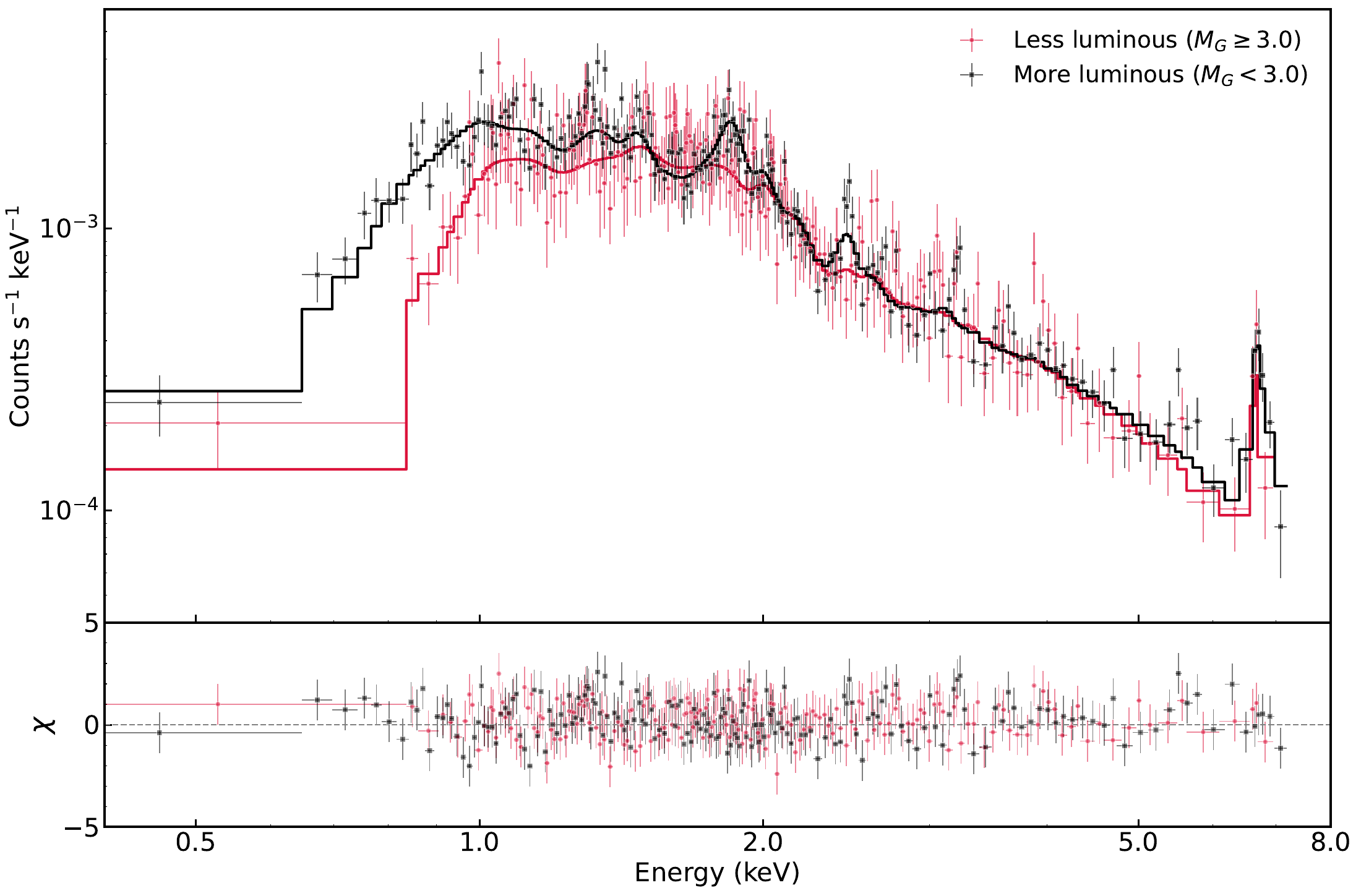}
  \caption{Stacked X-ray spectra of the SySt candidates divided into two subgroups based on their $M_G$. The black and red symbols represent the more-luminous ($M_G < 3.0$, 76 sources) and less-luminous ($M_G \geq 3.0$, 31 sources) samples, respectively. The solid lines denote the best-fit models, and the bottom panel shows the residuals in units of $\chi$. The consistent detection of a prominent 6.7 keV Fe line in both subgroups confirms that this high-ionization emission is a robust feature across different luminosity regimes.}
  \label{fig:LPV2spec}
\end{figure*}

\end{appendix}
\end{document}